\documentclass{aastex63}
\usepackage{graphicx}
\usepackage{amsmath}
\usepackage{epsfig}
\usepackage{bm}

\newcommand{\degree}{$^{\circ}~$}

\submitjournal{ApJ}
\shorttitle{HVCs in GC}
\shortauthors{Zhang et al.}

\begin{document}

\title{The Starburst Acceleration of High-Velocity Clouds in the Galactic Center}

\author[0000-0001-8261-3254]{Mengfei Zhang}
\affil{School of Physics, Zhejiang University, Hangzhou, Zhejiang 210023, China}
\email{murphychang@zju.edu.cn}

\author{Miao Li}
\affil{School of Physics, Zhejiang University, Hangzhou, Zhejiang 210023, China}
\email{miaoli@zju.edu.cn}

\author{Peixin Zhu}
\affil{School of Physics, Zhejiang University, Hangzhou, Zhejiang 210023, China}

\begin{abstract}
High-velocity clouds (HVCs) in the Galactic center have garnered significant attention due to their mysterious formation, potentially linked to starburst events or supermassive black hole activity in the region. However, it remains challenging to explain the observed column density and velocity distribution of HVCs. The discovery of high-velocity molecular clouds (HVMCs), which are denser and more massive, adds to this complexity. To address this, we conduct three-dimensional numerical simulations to explore the origin and magneto-hydrodynamic evolution of HVCs in the context of a starburst in the Galactic center. By incorporating magnetic fields and an initial tangential velocity for the clouds, our simulation results align with the observed properties of HVCs, supporting the notion that these clouds can originate from a starburst process. In addition, $\sim$ 5\% of the total mass of initial clouds can survive after 3.5 Myr, as a result, the following star formation will be more efficient than a feedback process that destroys all cool clouds.
\end{abstract}


\section{Introduction}\label{sec:intro}
Galactic winds, especially the nuclear winds, shape the gas distribution and metal content of galaxies, which are now commonly accepted as an important process affecting the galactic evolution \citep[e.g.][and references therein]{2012ARA&A..50..455F, 2014ARA&A..52..589H, 2017hsn..book.2431H, 2017ARA&A..55...59N, 2018Galax...6..114Z}. %
Due to its proximity, the center of the Milky Way provides a unique laboratory for studying the feedback process in detail, but its activity is relatively weak at present \citep{2003ApJ...591..891B, 2016A&A...589A..66H}. However, if ƒthe galactic wind was stronger in the past, it would produce some corresponding relics. It would be significant to detect these relics and study their relation with the galactic wind.

In the past few decades some such relics have been identified  at radio, X-ray and $\gamma$-ray bands, such as the Galactic Center Lobe (GCL; \citealp{1984Natur.310..568S}), the microwave haze \citep{2004ApJ...614..186F, 2013A&A...554A.139P}, the polarized lobes \citep{2013Natur.493...66C}, the Fermi bubbles \citep{2010ApJ...724.1044S}, the radio bubbles \citep{2019Natur.573..235H}, the X-ray chimneys \citep{2019Natur.567..347P} and the eROSITA bubbles \citep{Predehl2020}.
These structures have scales ranging from $\sim$100 pc to $\sim$10 kpc, suggesting that they originated from several distinct episodes of violent activity.
In addition, in the Galactic center, many high-velocity clouds (HVCs) were detected \citep{2004ApJ...605..216C, 2005ApJ...623..196C, 2018ApJ...855...33D, 2020ApJ...888...51L, 2020ApJ...898..128A}, and two high-velocity molecular clouds (HVMCs) are also discovered inside the HVCs \citep{2020Natur.584..364D}, but their origin has not been determined.
The HVCs are located at the transition region between the radio bubbles/X-ray chimney and the Fermi/eROSITA bubbles, therefore, the HVCs are possibly a bridge to connect the understanding of these relics at different scales.
In this case, HVCs possibly originate from a similar process of acceleration by the Galactic nuclear wind.
The local standard of rest (LSR) velocities of these HVCs can reach $\sim$ 300 km s$^{-1}$ at $\sim$ 2 kpc \citep{2020ApJ...888...51L}.
Based on an analytical biconical model, they should survive an acceleration process and have a lifetime of 4-10 Myr \citep{2020ApJ...888...51L}.
Nevertheless, the radiation and hydrodynamics mechanisms of their formation are still the subject of intense debate.

The formation of Fermi bubbles can be explained by the hadrons emission produced by the star formation \citep{2011PhRvL.106j1102C,Fujita2013,Fujita2014,Crocker2015} or leptons emission produced by Eddington outburst from the supermassive black hole, Sgr A* \citep{Zubovas2011, 2012MNRAS.424..666Z}.
These two kinds of models yield timescales of $\sim$ 8 Gyr and $\sim$ 6 Myr, respectively, which is due to the longer cooling time of hadrons compared to leptons. If the star formation rate is much higher, the leptons emission in a star formation model can also match the observation, and the timescale can decrease to 30 Myr \citep{Sarkar2015}. \citet{Guo2012,2013MNRAS.436.2734Y,Yang2017,Zhang2020,2022NatAs...6..584Y} propose an AGN jet model, which prefers a leptonic emission and a timescale of 1--6 Myr, while  \citet{Mou2014,Mou2015} suggest the bubbles are inflated by winds launched from the hot accretion flow in Sgr A* and the timescale is $\sim$ 10 Myr. In addition to the Fermi bubbles, some different possible relics have also been investigated. \citet{2024NatAs.tmp..228Z} analyze the multi-wavebands data, and suggest the eROSITA bubbles originate from the active star forming region, while \citet{2023NatCo..14..781M} suggest the eROSITA bubbles are produced by the circumgalactic medium wind. The radio bubbles at smaller scale may originate from multiple supernovae explosions \citep{2021ApJ...913...68Z}.

Some models can simultaneously explain different structures. The microwave "haze" and extended X-ray emission surrounding the Fermi bubbles may be reproduced by a star formation process \citep{2011PhRvL.106j1102C, Crocker2015}. This model can explain the location, luminosity and spectrum of radio, polarized radio lobe and $\gamma$-ray emission, and the extent, density and temperature of plasma in this region, but it is based on some rough assumptions, such as the cosmic ray and magnetic field distribution. \citet{Mou2014, Mou2015} suggest an AGN wind model to explain the relation between the radio/X-ray emission surrounding the Fermi bubbles, in which the decay of neutral pions generated in the collisions, combined with the inverse Compton scattering of background soft photons by the secondary leptons generated in the collisions and primary CRe, can well explain the observed $\gamma$-ray spectral energy distribution.
However, this model does not account for the formation of microwave "haze" and polarized lobe.
\citet{Sarkar2015} also propose a star formation model to explain the radio/microwave/X-ray/$\gamma$-ray emission, which can also explain some absorption clumps in similar region, but they apply a two-dimensional simulation and neglect the magnetic field.
The X-ray chimney possibly originates from the mini-starburst that produced the radio bubbles \citep{2021ApJ...913...68Z}, but the model can only explain these structures at small scales. The eROSITA/Fermi bubbles can also be inflated by an AGN jet \citep{2022NatAs...6..584Y}, though other structures are difficult to reproduce in this framework. Most of these models exhibit self-consistency, and subsequent simulations have provided further validation of their viability \citep[e.g.]{Guo2012, Mou2014, Mou2015, Sarkar2015, Zhang2020, 2021ApJ...913...68Z, 2022NatAs...6..584Y, 2024A&ARv..32....1S}. It is challenging to determine which model is correct, because the radiation features of starburst and AGN process are similar at large scale.

However, currently, there is no magnetohydrodynamic (MHD) simulation work to study the location and velocity population of HVCs in the Galactic center, thus the formation of HVCs and their relation with other relics are still ambiguous. The HVCs are expected to be gradually accelerated to high-velocity without disruption. For this to occur, the energy input driving the HVCs must be sufficiently strong to enable efficient acceleration, and gentle enough to ensure their survival. This paradox means the condition to produce HVCs is more stringent than other relics, making the study of HVCs a valuable tool for further constraining theoretical models. For example, multi-supernovae explosions can periodically produce the stable gentle wind, which is difficult to be produced by an AGN, since AGN usually release an amount of energy persistently or the period is unstable. If the relation between HVCs and other relics is robust, it could potentially be used to distinguish among different models for their origin.

A recent simulation work considered a nuclear wind naturally produced by supernova explosions to explain the formation of HVMCs \citep{2024MNRAS.527.3418Z}, suggesting that the HVCs can also survive supernova shock waves, as HVMCs are more challenging to form than HVCs. If the HVMCs can be produced by a starburst event, the HVCs may originate from similar process. At present, the Milky Way is quiescent, but a starburst event is still a possible explanation since the Galactic center exhibited a higher star formation rate about tens of millions of years ago  \citep{NoguerasLara2019}. Moreover, the clouds outflow is universal in active star-forming galaxies \citep{2018Sci...361.1016S, 2020MNRAS.493.3081R, 2020ApJ...905...86S, 2021A&A...653A.172S, 2023ApJ...944..134B}, which implies a similar outflow in the Milky Way is possible, if its activity is comparable to those star-forming galaxies.
All of these suggest that a starburst model is a possible solution for the formation of HVCs.

In this paper, we explore whether a starburst model can explain the properties of HVCs in the Galactic center, which can be taken as an expansion of \citet{2024MNRAS.527.3418Z}.
Unlike the previous study on high-velocity molecular clouds (HVMCs), this work employs a variety of configurations to better match the initial conditions of HVCs.
To investigate the distribution of HVCs over a larger region, the simulation box width has been expanded from 100 pc to 1500 pc. Correspondingly, the gravitational potential now incorporates more large-scale components, such as the halo and bulge, and the supernovae rate has been increased from $10\rm~kyr^{-1}$ to $20\rm~kyr^{-1}$.
The number of initial clouds has been increased from 1 to 50, and the density profile has been adjusted to more accurately reflect the Galactic center environment.
Furthermore, these clouds are initialized with rotational motion consistent with the Milky Way, in contrast to the static clouds used in the HVMCs study.
The most reliable observational properties of HVCs--their spatial and velocity distributions--can be directly derived from the simulation results and serve as critical benchmarks to assess the validity of the simulations. Additionally, the simulated emission in the X-ray and optical wavebands can be compared with observational data to further validate the model.


\section{Simulation}\label{sec:sim}

The publicly available, modular magnetohydrodynamic  code \textit{PLUTO}\footnote{http://plutocode.ph.unito.it/} \citep{Mignone2007, Mignone2012} is utilized in this work to perform the simulations.
The code is the same as the work of \citet{2024MNRAS.527.3418Z}, but the version is updated from \textit{PLUTO 4.3} to \textit{PLUTO 4.4}.
The changes between different versions will not significantly affect the simulation results.
The simulation configuration is based on the work of \citet{2024MNRAS.527.3418Z}, with several key parameters adjusted to more accurately reflect the observed properties of HVCs.

\subsection{Basic configuration}\label{subsec:config}
{This grid-based MHD code \textit{PLUTO} employs a second-order Runge–Kutta time integrator and a Harten-Lax-van Leer Riemann solver for middle contact discontinuities, which is compatible to the high-resolution shock-capturing (HRSC) schemes \citep{Mignone2007}. Comparing with traditional finite difference codes, the motivation behind such schemes is the ability to model strongly supersonic flows while retaining robustness and stability. Implementation of HRSC algorithms is based on a conservative formulation of the fluid equations, and proper upwinding requires an exact or approximate solution to the Riemann problem, i.e., the decay of a discontinuity separating two constant states. Most HRSC algorithms are based on the  reconstruct-solve-average strategy. In this approach, volume averages are first reconstructed using piecewise monotonic interpolants inside each computational cell. A Riemann problem is then solved at each interface with discontinuous left and right states, and the solution is finally evolved in time. Such a scheme is particularly suitable for time-dependent, explicit computations of highly supersonic flows in the presence of strong discontinuities, making it well-suited for simulating the interaction between the supernova shock and the molecular clouds.}

The simulation is based on a three-dimensional (3D) MHD Cartesian frame with a grid of $\rm 750 \times 750 \times 500$, equivalent to a physical volume of $\rm 1500 \times 1500 \times 1000$~pc$^3$ and a linear resolution of 2 pc cell$^{-1}$.
To ensure the reliability of the results at lower resolution, phase diagrams generated at low resolution are compared with those at high resolution. For instance, the fiducial case of \citet{2024MNRAS.527.3418Z} was tested at lower resolution, and the resulting phase diagrams showed no significant qualitative differences, allowing the lower-resolution case to be expanded to a wider simulation box.
The simulation box is smaller than the survey region of HVCs observation \citep{2018ApJ...855...33D, 2020ApJ...888...51L}, but can contain most of HVCs and their statistical features.
Only the northern part of  the Milky Way is considered due to its symmetric distribution.
The $z$-axis is set to be perpendicular to the Galactic disk (north as positive), the $y$-axis to run along decreasing Galactic longitude, and the $x$-axis to be parallel to the line-of-sight (the observer at the positive side).
An outflow boundary condition is adopted for all directions, which means that some material may flow outside the simulation box.
In this work, the distance of Galactic center is set to be 8.5 kpc, consistent with the observation work of HVCs \citep{2013ApJ...770L...4M}.

The simulation is governed by the ideal MHD conservation equations,
\begin{eqnarray}
      \begin{cases}
      \dfrac{\partial \rho}{\partial t} + \nabla \cdot (\rho \bf{v}) = 0,\\
      \\
      \dfrac{\partial (\rho\bf{v})}{\partial t}+\nabla \cdot\left[\rho\bf{vv}+\bf{1}p\right]^{T}=-\rho \nabla \Phi, \\
      \\
      \dfrac{\partial E_{t}}{\partial t}+\nabla \cdot\left[\left(\dfrac{\rho \bf{v}^{2}}{2}+\rho \epsilon+p+\rho \Phi\right) \bf{v}- \dfrac{\bf{v} \times \bf{B} \times \bf{B}}{4\pi}\right] = -\dfrac{\partial\left( \rho \Phi\right)}{\partial t}, \\
      \\
      \dfrac{\partial \bf{B}}{\partial t} - \nabla \times (\bf{v} \times \bf{B}) = 0,
      \end{cases}
\end{eqnarray}
where $\rho$ is the mass density, $p$ the thermal pressure, $\bf{v}$ the velocity, $\bf{B}$ the magnetic field, $\bf{1}$ the dyadic tensor, $\Phi$ the gravitational potential, and $E_t$ the total energy density, defined as:
\begin{eqnarray}
  E_t = \rho \epsilon + \frac{(\rho\bf{v})^2}{2\rho} + \frac{\bf{B}^2}{8\pi},
\end{eqnarray}
where $\epsilon$ is the internal energy.
We use an ideal equation of state, i.e., $\epsilon = p/ (\Gamma -1)$, in which the ratio of specific heats is $\Gamma$ = 5/3.

The simulation applies a new Galactic gravitational potential model \citep{2023A&A...680A..40M}, which includes the contribution of stellar disk (thin and thick disk), gaseous components (HI and H$_2$), bulge and halo (see Appendix~\ref{sec:grav}), therefore, this reflects a more realistic gravity than the work of \citet{2024MNRAS.527.3418Z}, which only considers the supermassive black hole, the nuclear star cluster, and the nuclear disk.
The radiative cooling in the simulation is described by a piecewise cooling function (see Appendix~\ref{sec:cf}), with a cooling lower limit of 100 K and a solar abundance (H abundance $X_{\odot}$=0.711, He abundance $Y_{\odot}$=0.2741, metallicity $Z_{\odot}$=0.0149) for the initial molecular clouds and interstellar medium (ISM).
The multiphase gas in the Galactic center includes hot ionized ($\rm \sim 10^6$ K) \citep{2013ApJ...779...57K,2019Natur.567..347P}, warm ionized (10$^4$ to 10$^5$ K) \citep{2015ApJ...799L...7F, Bordoloi2017} and cool atomic (10$^3$ to 10$^4$ K) gas \citep{2013ApJ...770L...4M, 2018ApJ...855...33D}, in which the gas below 100 K is usually taken as molecular gas and has a weak cooling process.

\subsection{Supernova explosion and gaseous clouds}\label{subsec:sw}
In the simulation, the supernovae randomly explode in a cylindrical region centered at Sgr A*, where the height and radius are respectively set to be 10 pc and 150 pc, similar to the scale of the central molecular zone and nuclear disk.
The supernova birth rate is $20\rm~kyr^{-1}$, which is estimated by assuming a star formation rate (SFR) of 2 M$_{\odot}$ yr$^{-1}$, an initial mass function (IMF) \citep{Kroupa2001} and a minimum mass of 8 M$_{\odot}$ for the progenitor star of a core-collapse supernova \citep{2018ApJ...863..127K}.
Currently, the SFR in the Galactic center is 0.1 M$_{\odot}$ yr$^{-1}$ \citep{Barnes2017,2020MNRAS.497.5024S}, while the average SFR can increase to $0.2-0.8\rm~M_{\odot}~yr^{-1}$  in the past 30 Myr \citep{NoguerasLara2019}.
Actually, the SFR can increase in a short amount of time so that it can go to 2 M$_{\odot}$ yr$^{-1}$ within few millions of years, consistent with the value used in the simulation.

The way to simulate one supernova explosion is the same as \citet{2024MNRAS.527.3418Z}. The methodology follows the work of \citet{Truelove1999}, configuring initial density, velocity, and temperature profiles within a generated zone with an initial radius of 5 pc. We rigorously validated the initial simulated evolution by comparing the results with the analytical calculations provided by \citet{Leahy2017a}. Each supernova was initialized with a typical explosion energy and mass of 10$^{51}$ erg and 10 M$_{\odot}$ \citep{Sukhbold2016}, respectively, in which the kinetic energy constitutes the majority of the explosion energy. In this simulation, the galactic wind is a direct outcome of the collective energy from these supernovae. The shock wave from different supernovae will interact and mix with each other, producing a large shock region and forming the galactic wind.

Initially, to match the total mass of observed HVCs, 50 clouds are set to distribute in a cylindrical region with a height of 50 pc and a radius of 150 pc.
Such a region is possible for clouds to reach based on recent observations, in which some clouds are located between 15 $\sim$ 55 pc above the Galactic plane \citep{2024A&A...689A.121V}.
The density profile of these clouds is set to be $\rm n_c = (n_0-n_{ISM})/(1+r^2) + n_{ISM}$ \citep{BandaBarragan2016}, in which n$_0$ is the central density, n$_{ISM}$ the ISM density, r the radius.
In the fiducial simulation, n$_0$ = 1500$\sim$2000 H cm$^{-3}$, n$_{ISM}$ = 0.01 H cm$^{-3}$ and the maximum radius of the initial clouds is 10 pc.
The total mass of these clouds is $\sim$ 130000 M$_{\odot}$, comparable to the observed total mass of HVCs above the Galactic plane.
The initial tangential velocity of these clouds is set to balance the local gravitational potential.

In addition, once injected, the ejecta of supernovae will eventually partially mix with the clouds, and change their metallicity.
To study the mixture, we introduce two tracer parameters, $Q_1$ and $Q_2$, which are both evaluated at each cell and follow a simple conservation law:
\begin{eqnarray}
      \frac{\partial (\rho Q_i)}{\partial t} + \nabla \cdot (\rho Q_i \bf{v}) = 0.
\label{eqn:tracer}
\end{eqnarray}
$Q_1$ has a value of 1 for pure supernova ejecta and 0 for other components while
$Q_2$ has a value of 1 for pure molecular clouds and 0 for other components.
The values in between indicate a mixed gas.
These tracer parameters allow us to track the mixing process over time and analyze the distribution of metals.

\subsection{The ISM and the magnetic field}\label{subsec:runs}
The simulation is initialized with a uniform distribution of ISM density and temperature, with values of 0.01 H cm$^{-3}$ and 10$^6$ K, respectively, over the entire simulation box. There are usually two choices for ISM distribution, uniform ISM and hydrostatic equilibrium ISM. Although thermal pressure is expected to be higher at lower latitudes due to rough hydrostatic equilibrium against gravity, our preliminary tests suggest that this effect is unimportant since the shock wave from the supernovae breaks this equilibrium early on.
Moreover, the stellar wind in the Galactic center is also strong and can unremittingly break this equilibrium.
The distribution of a hydrostatic equilibrium ISM is not necessary.

The magnetic fields in the Galactic center include many different components, particularly in the central tens of parsecs \citep{Ferriere2009}, which makes it a challenging problem.
There is actually a general model for the whole Milky Way \citep{2013lsmf.book..215B, 2017JCAP...10..019C}, in which the magnetic field is parallel to the Galactic plane at lower latitude and gradually tends to be perpendicular at higher latitude and more central regions, but this model is not accurate in the Galactic center.
The magnetic strength ranges from $\sim$ 1 mG in the central tens of parsecs \citep{Ferriere2009} to few $\mu$G at 1 kpc above the Galactic plane \citep{2017JCAP...10..019C}.
For simplicity, we in this work adopt a perpendicular magnetic direction and a homogeneous magnetic strength of 10 $\mu$G over the whole simulation box, similar to the fiducial run of \citet{2024MNRAS.527.3418Z}.

\section{Results and Discussion}\label{sec:res}

To identify HVCs in the simulation, we adopt a strict criterion: an HVC is defined as a cloud composed of more than ten connected cells, with each cell having a number density greater than 1 H cm$^{-3}$ and a temperature below 10$^3$ K. Given the simulation's linear resolution of 2 pc per cell, ten cells correspond to a volume of 80 pc$^{3}$, which is sufficiently large to ensure it is not an artificial spot.
Because we set a lower limit for the cooling temperature, the temperatures below 100 K are typically not due to radiative cooling, but rather due to adiabatic expansion.
It is worth noting that some atomic components may survive under conditions of lower density and higher temperature. As a result, this strict criterion may lead to an underestimation of both the number and total mass of HVCs.

Based on the observations, we can derive many parameters, such as the mass, volume density and cloud radius of HVCs, but they are usually dependent on some assumptions.
For example, to estimate the mass of a cloud from the column density, a regular shape is assumed for the cloud, then the density and radius can also be derived. However, this assumption of regular shape is obviously unreliable.
Therefore, we take the location and velocity distribution as the main criteria for comparing the simulation results with observations, while treating those derived parameters as supplementary references.
The aim of this work is to simultaneously reproduce the acceleration and survival of single cloud, and the statistical column density and velocity distribution of multiple clouds.

\subsection{Column Density and Velocity of HVCs in the simulation}\label{subsec:cv}

\begin{figure}
\includegraphics[width=1\textwidth]{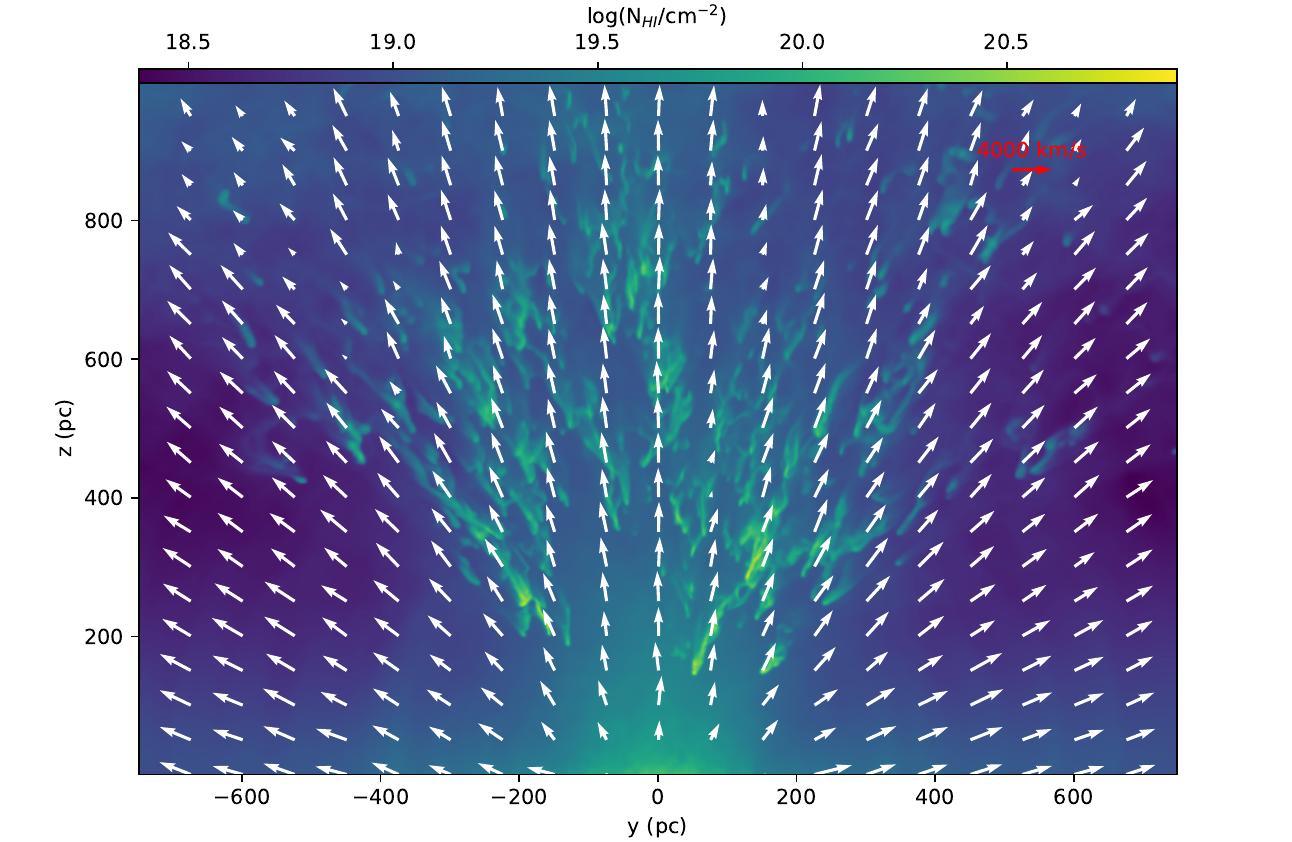}
\includegraphics[width=1\textwidth]{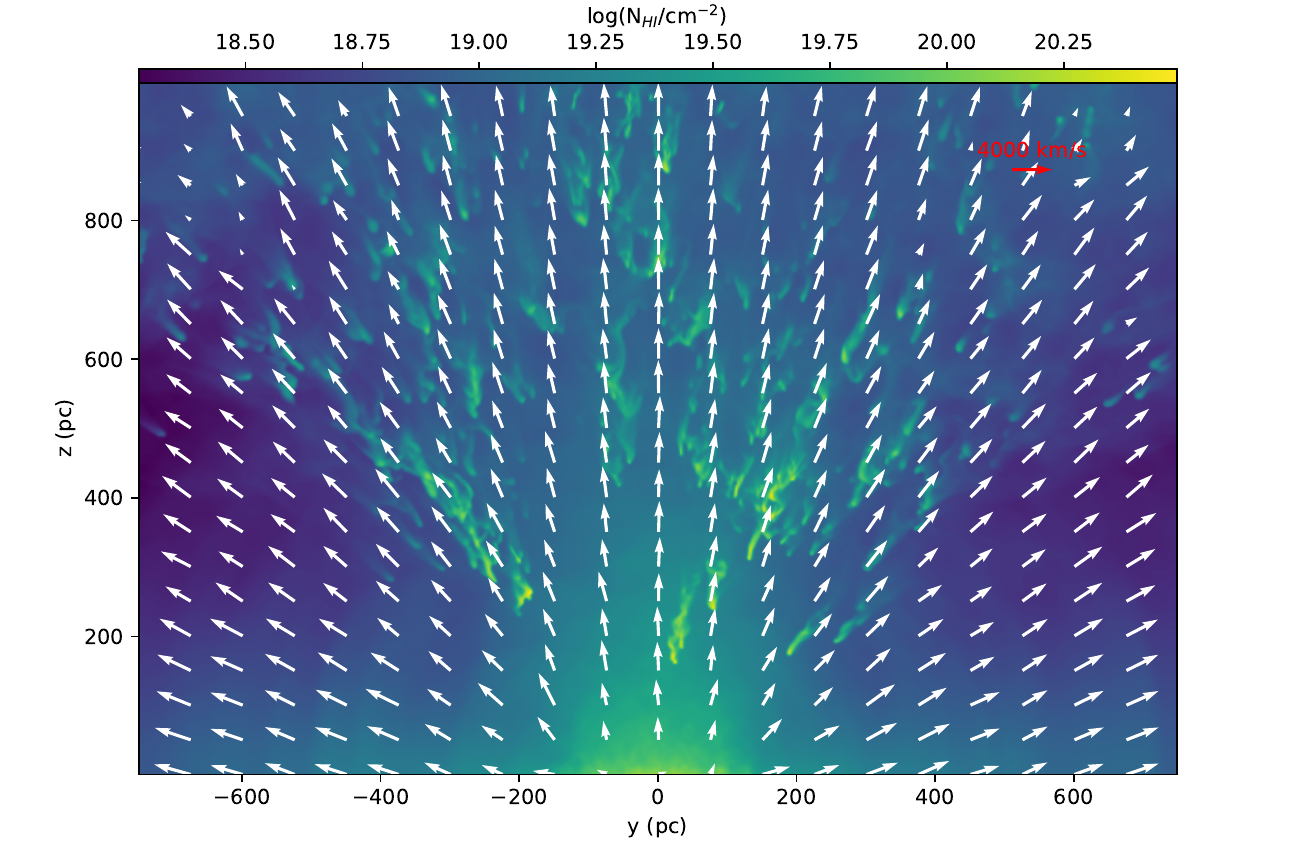}
\caption{The column density and velocity distribution after 3 and 3.5 Myr. The colorful patterns show the column density integrated along $x$-axis, and the white arrows show the velocity at $x$=0. The red arrows show the length of 2000 km s$^{-1}$.
\label{fig:sim}}
\end{figure}

\begin{figure}
\includegraphics[width=0.5\textwidth]{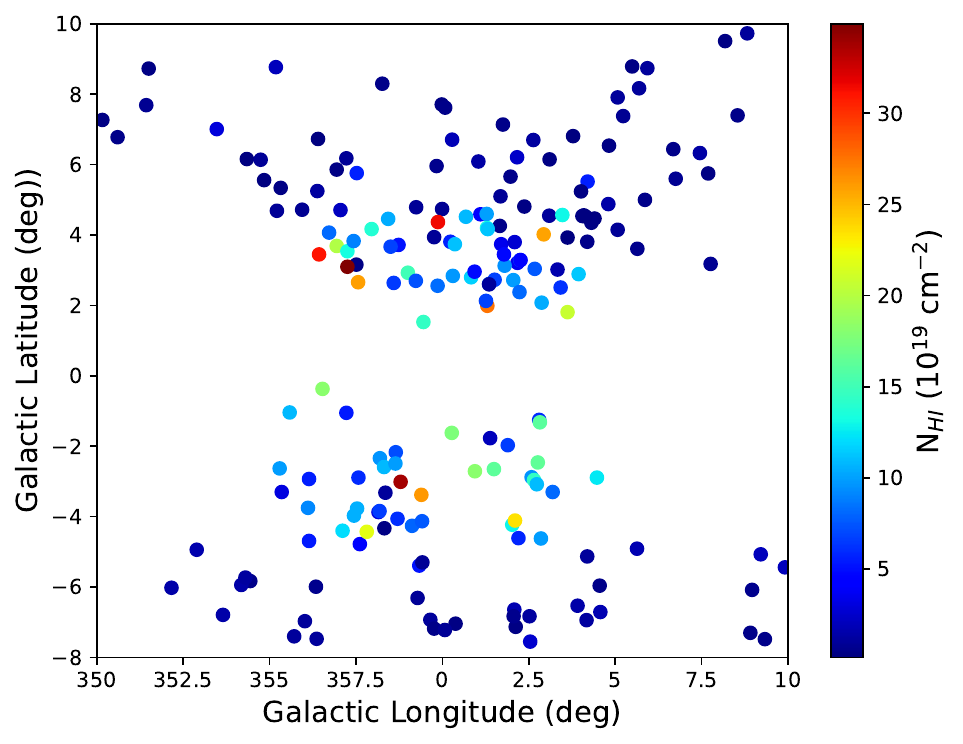}
\includegraphics[width=0.5\textwidth]{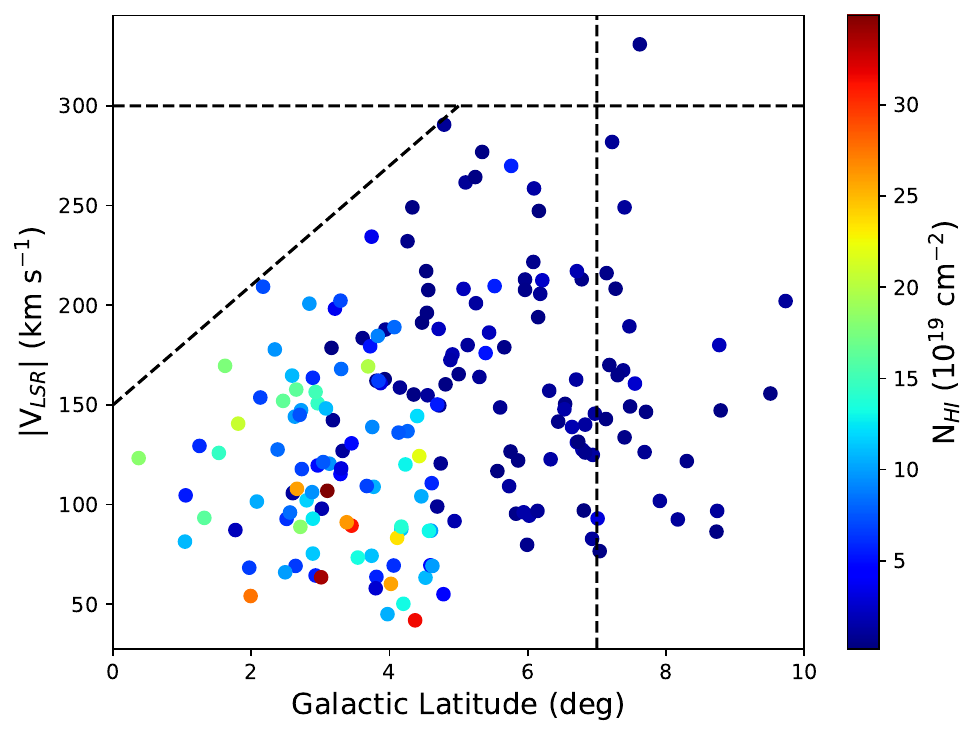}
\includegraphics[width=0.5\textwidth]{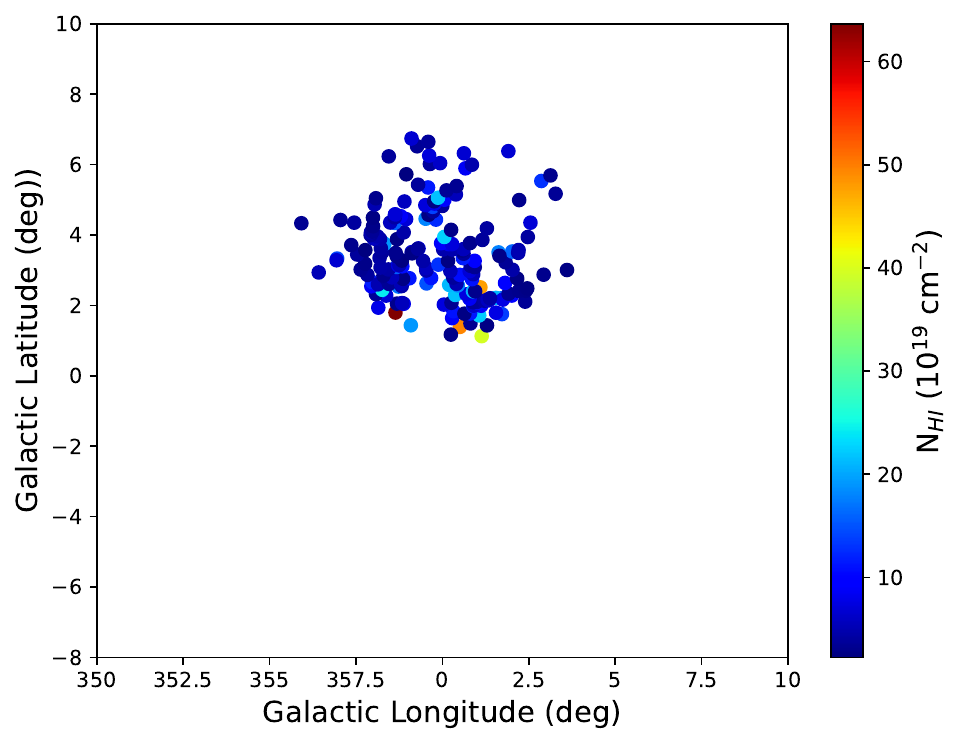}
\includegraphics[width=0.5\textwidth]{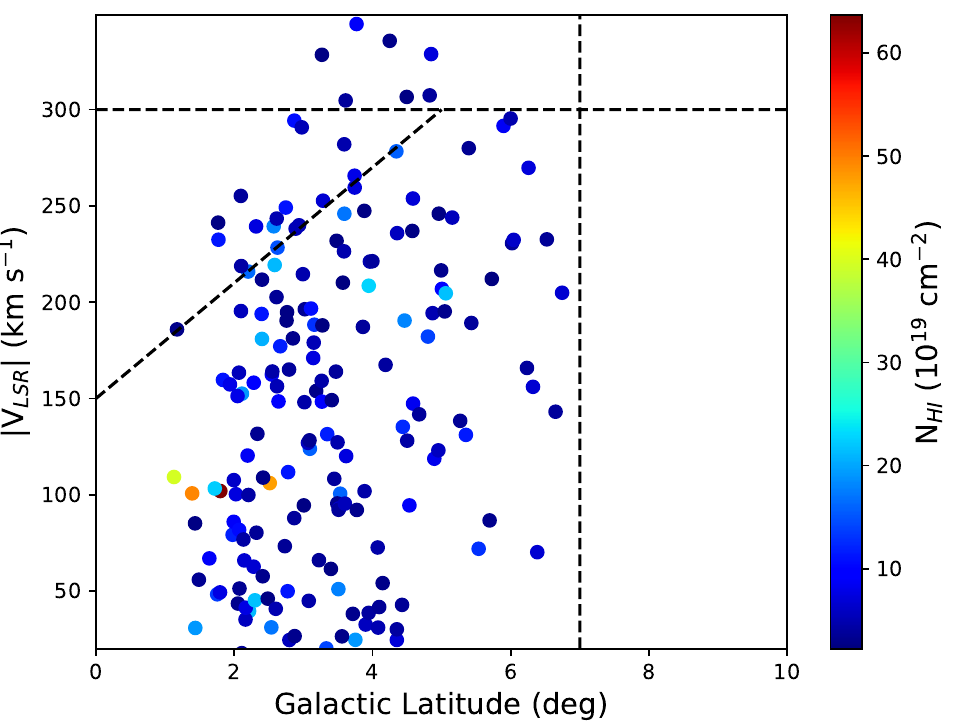}
\includegraphics[width=0.5\textwidth]{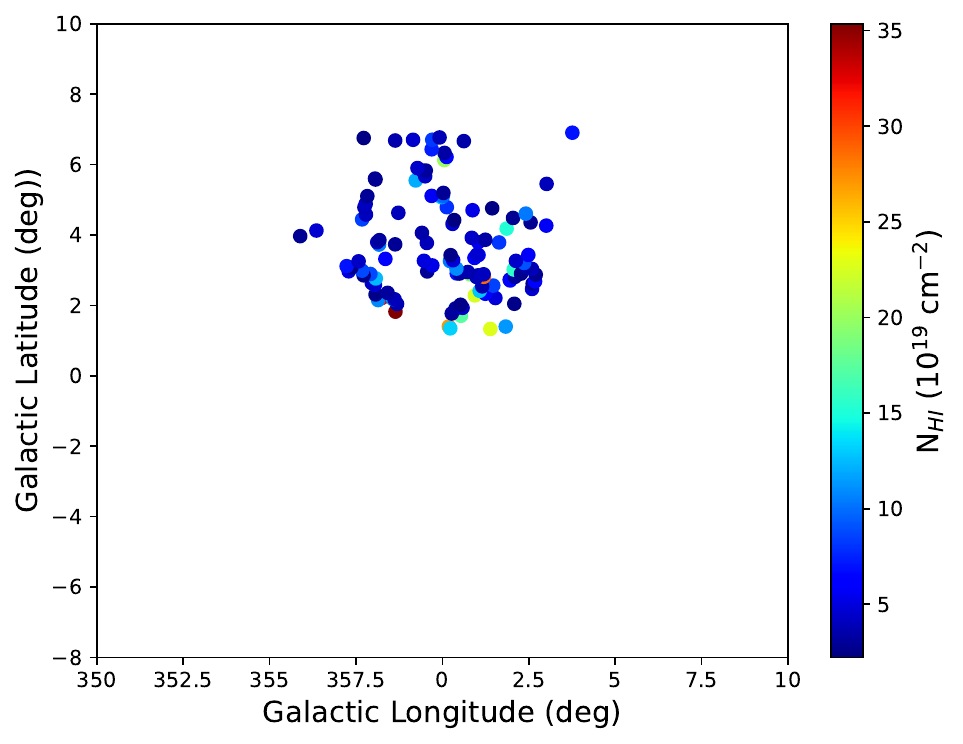}
\includegraphics[width=0.5\textwidth]{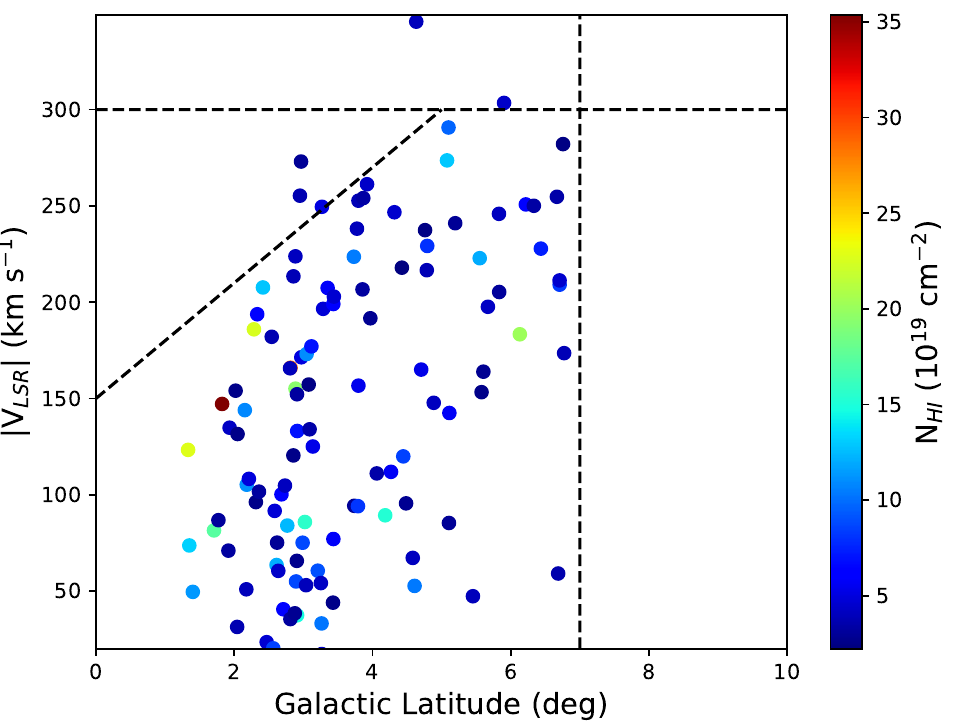}
\caption{\textit{Left}: the location distribution of HVCs. From top to bottom, the figures show the location and column density of HVCs in observation, and in simulation after 3 and 3.5 Myr. \textit{Right}: the absolute velocity distribution of HVCs along Galactic latitude. From top to bottom, the figures show the velocity and column density of HVCs in observation, and in simulation after 3 and 3.5 Myr. The horizontal and perpendicular dotted lines show the velocity of 300 $\rm km~s^{-1}$ and Galactic latitude of 7\degree, respectively. The oblique line shows the boundary of velocity distribution.
\label{fig:stat}}
\end{figure}

Figure~\ref{fig:sim} illustrates the column density and velocity, showing a similarity between the simulated HVCs and observations.
To better compare with the observation, we plot the simulated and observed location distribution of HVCs in the left panel of Figure~\ref{fig:stat}. The observational data are extracted from \citet{2018ApJ...855...33D} and \citet{2020ApJ...888...51L}, in which the HVCs typically exhibit a column density of 10$^{19}$ $\sim$ 10$^{20}$ cm$^{-2}$ \citep{2018ApJ...855...33D} and a outflow velocity of 50$\sim$300 $\rm km~s^{-1}$ \citep{2020ApJ...888...51L}.
The oblique line shows an upper velocity limit of HVCs at different latitudes.
After 3 Myr, most HVCs remain within the simulation box, but some begin to flow outside the box after 3.5 Myr.
The simulation box is smaller than the survey region of observations and many HVCs have been observed at distances outside our simulation box, therefore, the case after 3.5 Myr is more consistent with the observation.
Initially, the shock wave travels at speeds exceeding 4000 $\rm km~s^{-1}$, but its velocity gradually decreases due to gravitational forces and interactions with clouds and ISM.
Some clouds are disrupted, forming filaments and head-tail structures.
Most of the clouds will propagate to high latitude, but some material will be pulled back to the Galactic disk to form a high-density base visible in the figure.

The simulated open angle of HVCs is smaller than the observed $\sim$ 140\degree at larger scale \citep{2018ApJ...855...33D}, because this angle will decrease at lower latitude and the latitude range in the simulation is smaller than observations.
The radius of supernova explosions region is the same as the region of initial clouds distribution, causing most HVCs to propagate away from the Galactic plane rather than parallel to it.
Consequently, the number of HVCs at latitudes below 1\degree is significantly lower in the simulation.
In reality, the density in the Galactic disk is higher, and during a starburst, additional material may be pulled into a region with the central region. This would likely result in more HVCs at lower latitudes, increasing the fraction with velocity components parallel to the Galactic plane, thereby producing a larger opening angle.
Furthermore, the magnetic field in the Galactic center should consist of both vertical and horizontal components. The horizontal components would contribute to higher horizontal velocity components for some material \citep{2024MNRAS.527.3418Z}. In the absence of horizontal magnetic fields in the simulation, the resulting open angle is artificially reduced.

Figure~\ref{fig:sim} only shows the velocity at $y$-$z$ plane,  making it necessary to derive the LSR velocity for a more direct comparison with observational data.
To get the LSR velocity of HVCs, we firstly need to define HVCs in the simulation.
The total velocity of each HVC is estimated by a mass-weighted velocity:
\begin{equation}
       < v > = \dfrac{\int\ v \rho\ dV}{\int\ \rho\ dV},
\label{eqn:v}
\end{equation}
in which $v$ and $\rho$ are the velocity and density of each cell of one HVC.
Then the LSR velocity can be derived from the total velocity in terms of the classical method  (see Appendix~\ref{sec:lsr}).

The statistical LSR velocity along the $z$-axis is shown in the right part of Figure~\ref{fig:stat}, in which the velocity distribution after 3.5 Myr is more consistent with the observation.
The overall upper limit of the LSR velocity is 300 km s$^{-1}$, and the upper limit at low latitude follows an oblique line starting at 150 km s$^{-1}$ and increasing to 300 km s$^{-1}$.
The column density ranges from 2 $\times$ 10$^{19}$ cm$^{-2}$ to 3.5 $\times$ 10$^{20}$ cm$^{-2}$.
Notably, HVCs with higher column densities are typically found at lower latitudes and exhibit lower LSR velocities.
These features closely match the upper right panel of Figure~\ref{fig:stat}.

The confirmed HVCs is less than those shown in the top panel of Figure~\ref{fig:stat}, because this simulation only considers the region above the Galactic plane.
Additionally, the use of a strict criterion for identifying HVCs could contribute to this difference.
At the latitude below 2\degree, comparing with the observation, the HVCs is more difficult to reach a high velocity in the simulation, because the simulation does not account for the presence of outer, high-density gas in the Galactic disk, which could also be accelerated to high LSR velocities.
Furthermore, some of the observed low-latitude HVCs might not originate from the nuclear wind but instead result from Galactic rotation, which could also explain part of the discrepancy between the simulation and observational data.

The terminal velocity of the nuclear wind can reach 2000 km s$^{-1}$ at $z$ = 1 kpc, and the velocity decrease is slow at higher latitude due to the low density.
If we take 2000 km s$^{-1}$ as the mean velocity, the earliest wind front can reach 7 kpc after 3.5 Myr, almost the scale of eROSITA/Fermi bubbles.
Moreover, after 3.5 Myr, the energy input from supernovae is 7$\times~10^{55}$ erg, comparable to the total energy content of the Fermi bubbles, $\sim 10^{56}$ erg \citep{2013Natur.493...66C}, which implies the possible relation between these bubbles and HVCs.
In the simulation box, the total mass of confirmed HVCs after 3.5 Myr is $\sim$ 7000 M$_{\odot}$, i.e., $\sim$ 5\% of the total mass ($\sim$ 130000 M$_{\odot}$) of initial clouds, which is a significant fraction and provides valuable insights into the role of HVCs in the recycling of gas within the galactic ecosystem.
It should be noted that some clouds flow outside the simulation box after 3.5 Myr and some clouds material are excluded due to the strict criterion, therefore the actual amount of surviving HVCs is likely higher.
These surviving clouds are expected to act as "seeds" for the formation of giant molecular clouds in the next cycle, contributing to the regulation of SFR in the galaxy.

\citet{2017MNRAS.468.4801Z} claim the cool gas with temperature of 10$^2$ $\sim$ 10$^4$ K cannot survive a hot Galactic wind, but their study does not account for the effects of magnetic fields and primarily focuses on processes occurring at larger scales. .
A more detailed work that incorporates magnetic fields at a similar scale suggests that cool gas can indeed survive in such an environment\citep{2024MNRAS.527.3418Z}.
Overall, the proposed formation mechanism is plausible and can contribute to our comprehension of the enigmatic ecosystem in the Galactic center.

\subsection{Analysis on the X-ray and optical  emission}\label{subsec:X}
\begin{figure}
\includegraphics[width=0.9\textwidth]{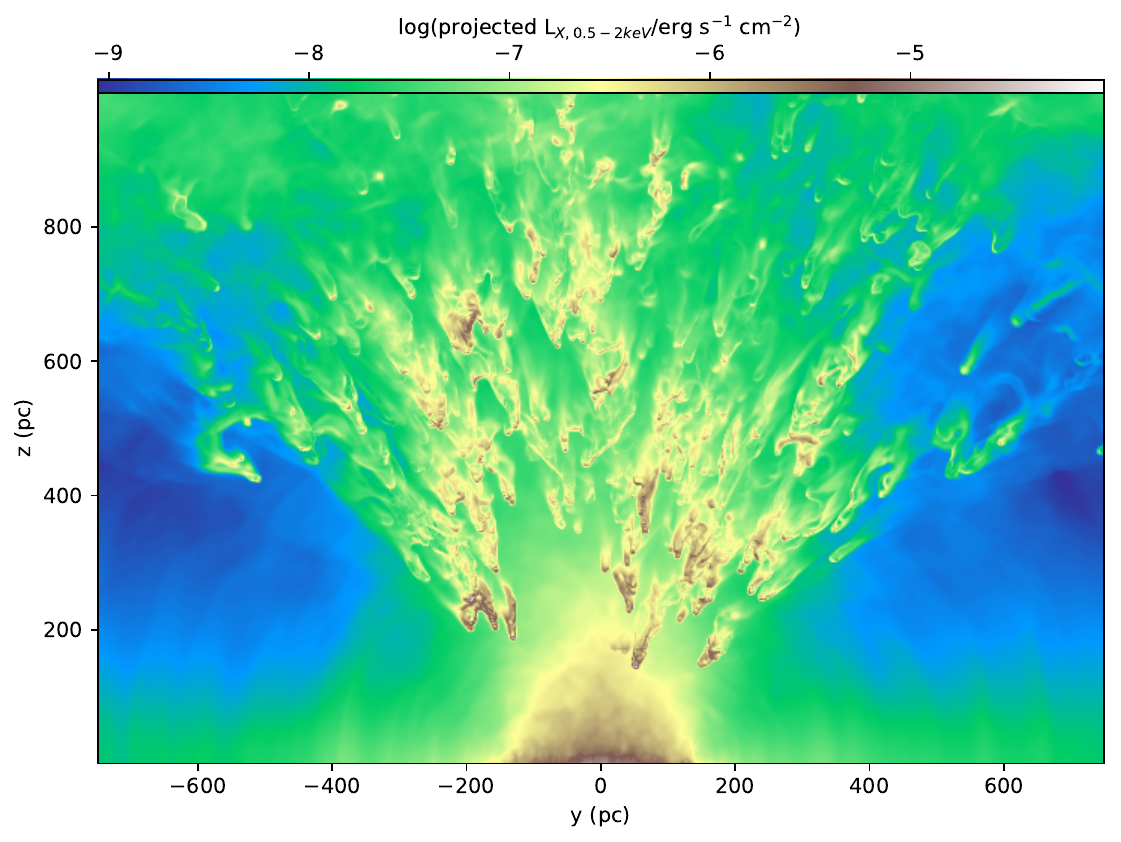}
\includegraphics[width=0.9\textwidth]{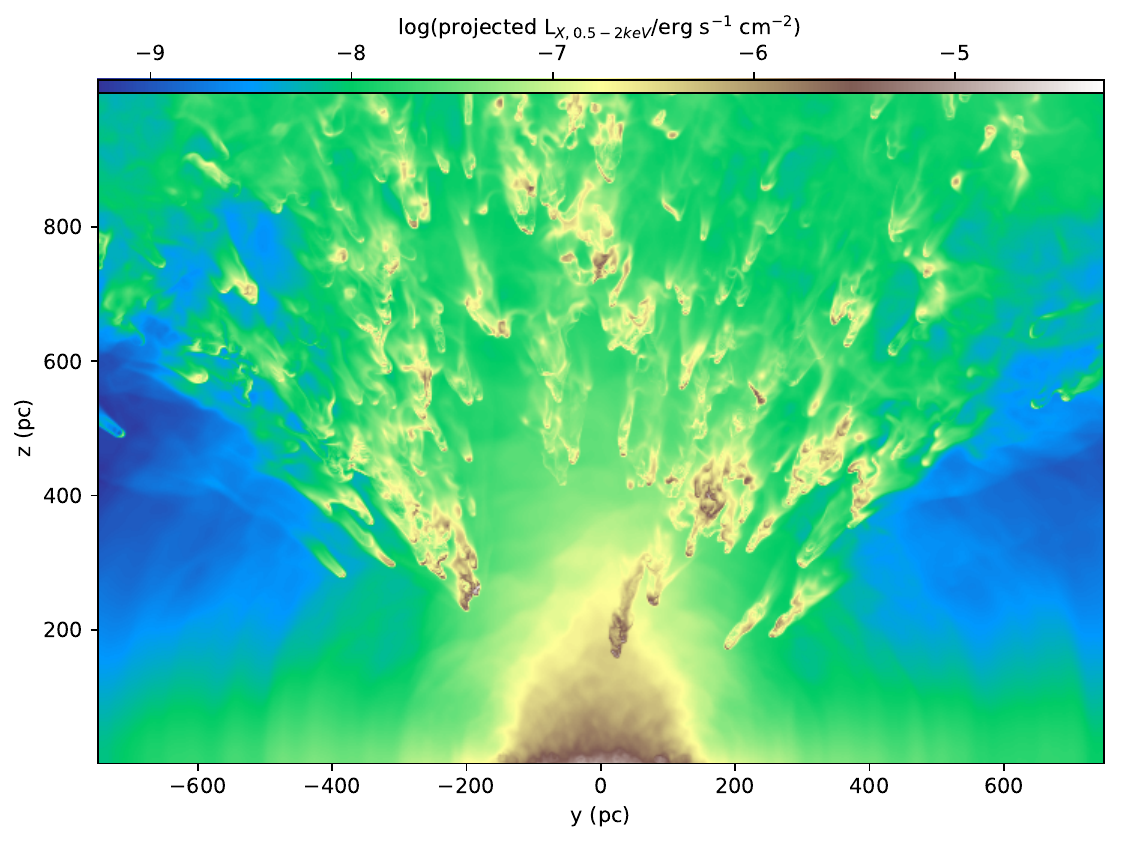}
\caption{The X-ray emission integrated along $x$-axis after 3 and 3.5 Myr.
\label{fig:X}}
\end{figure}

Figure~\ref{fig:X} shows the integrated X-ray luminosity at 0.5 -- 2 keV, which is calculated using  the database of \textit{ATOMDB}\footnote{http://www.atomdb.org}.
The shock wave from supernovae can strip and heat the surface gas of HVCs, then the stripped gas are ionized and produce X-ray emission.
At low latitudes, the forces acting on the material are a combination of the radial wind pressure and vertical gravity. The vertical gravity dominates, pulling some material back toward the Galactic disk. This process results in enhanced X-ray luminosity in the disk region, as well as the formation of hazy semicircular shells centered on the Galactic center.

The gas surrounding the HVCs appears brighter than the other ISM, leading to an inhomogeneous X-ray emission distribution, a pattern resembling the eROSITA bubbles observed at low latitudes \citep{Predehl2020}.
The pattern after 3 Myr shows some stripped gas of HVCs may already reach much higher latitude, even though most HVCs have not yet extended beyond 1 kpc.
After 3.5 Myr, some HVCs begin to flow outside the simulation box, potentially contributing to the X-ray emission at the base of the eROSITA bubbles.
The inhomogeneous X-ray emission within the eROSITA bubbles can be attributed to the uneven distribution of the ISM and the varying energy input. These HVCs could be one of the key factors behind this variability. However, while HVCs may be associated with X-ray emission, it is difficult to definitively confirm a causal link between the two diffuse sources solely based on their spatial correlation. Higher-resolution X-ray observations of specific compact HVCs will be essential for a deeper understanding of this connection.
The shock wave can also accelerate particles to be cosmic rays, then produce X-ray and $\gamma$-ray emission, which can possibly be used to explain the relation between eROSITA bubbles and Fermi bubbles\citep{2010ApJ...724.1044S}.

In addition to the large-scale eROSITA/Fermi bubbles, smaller-scale structures such as the X-ray chimneys and radio bubbles, with heights of approximately 200 pc above the Galactic plane, were detected by \citet{2019Natur.567..347P,2019Natur.573..235H}.These features are located at the base of the HVCs discussed in this study. The inferred age of the HVCs from our simulations is on the order of a few million years, which is roughly consistent with the dynamical timescale of the radio bubbles and the Fermi bubbles originally suggested by \citet{2019Natur.573..235H} and \citet{2013MNRAS.436.2734Y}, indicating a potential connection.
However, their timescales actually have not been resolved, the radio bubbles may be younger \citep[330 kyr]{2021ApJ...913...68Z} and the Fermi bubbles may be much older \citep[1 Gyr]{2011PhRvL.106j1102C}.
In the context of the supernova-based model for the origin of the radio bubbles/chimneys \citep{2021ApJ...913...68Z}, the radio bubbles would be a dynamically younger and independent structure simply evolving in the interior of the Fermi/eROSITA bubbles, which themselves were formed by older activities in the Galactic center.
The phenomenon can be explained by a simplified model\citep{2024MNRAS.527.3418Z}, in which the star formation in the Galactic center has been episodic on a timescale of $\sim$10 Myrs \citep{2015MNRAS.453..739K}, then these relics and HVCs are established episodically by consecutive generations of mini-starbursts and collapses.
This pattern offers a comprehensive explanation for the interrelation between various feedback remnants, without necessitating the introduction of new mechanism.

\begin{figure}
\includegraphics[width=0.9\textwidth]{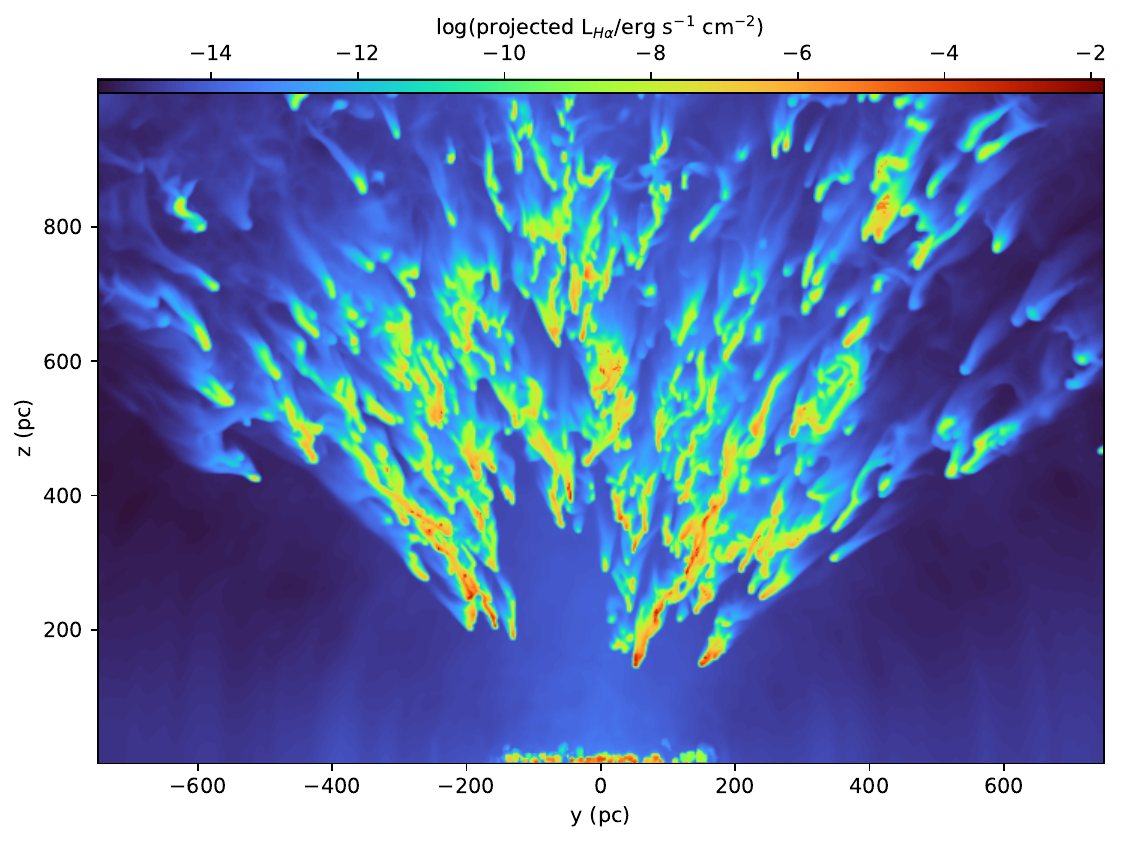}
\includegraphics[width=0.9\textwidth]{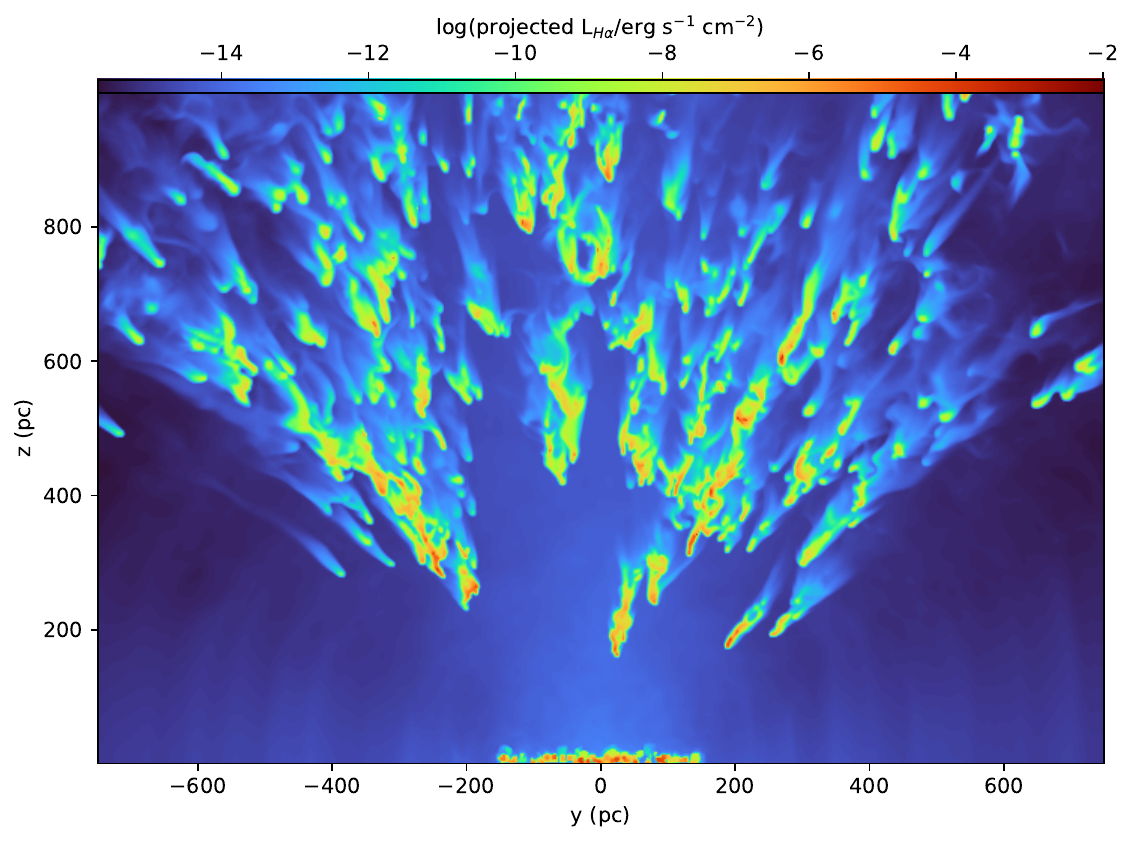}
\caption{The H$_{\alpha}$ emission integrated along $x$-axis after 3 and 3.5 Myr.
\label{fig:H}}
\end{figure}

Similar model has been proposed to explain the outflow of starburst galaxies \citep{2008ApJ...674..157C}, in which the X-ray emission is also suggested to originate from stripped gas of HVCs, meanwhile, the model shows a tight relation between soft X-ray emission and H$_{\alpha}$ emission.
Figure~\ref{fig:H} shows the H$_{\alpha}$ emission derived from our simulations, in which the correlation is also conspicuous by comparing it with Figure~\ref{fig:X}.
The X-ray emission is more diffusive than H$_{\alpha}$ emission, consistent with the work of \citet{2008ApJ...674..157C}.
Such a relation also implies the HVCs detected in the Galactic center have similar origin of those in starburst galaxies.

\section{Summary}\label{sec:sum}

To investigate the formation of HVCs in our Galactic center, we perform simulations based on a starburst model, where HVCs originate from low-latitude gaseous clouds accelerated by a subsequent supernova explosions.
The resultant density and velocity distribution of HVCs are well consistent with the observation, thus the proposed formation mechanism is plausible.
The following X-ray analysis predicts the X-ray emission associated with the HVCs and relates the HVCs and other possible feedback relics, based on which a long-duration model is suggested to simultaneously explain these structures.

Due to the limited size of the simulation box, the subsequent evolution of HVCs beyond 1 kpc latitude remains uncertain, but it is clear that they will propagate to higher latitude. Furthermore, the small box size restricts us to estimating the accurate total mass of clouds and studying the infalling gas to the disk at larger longitude. Moreover, the evolution of supernova shocks may be also different from the realistic condition. Future efforts involve expanding the simulation box, simplifying supernova explosion settings, and implementing adaptive mesh refinement to provide a more comprehensive understanding of the phenomena.

\section*{Acknowledgements}
We acknowledge the cosmology simulation database (CSD) in the National Basic Science Data Center (NBSDC) and its funds the NBSDC-DB-10. We acknowledge the support from the National Key Research and Development Program of China (2022YFA1602903), from the National Natural Science Foundation of China (12147103, 12273010), and from the Fundamental Research Funds for the Central Universities(226-2022-00216).

\section*{Data Availability}
The simulation code and data underlying this article may be shared upon reasonable request to the corresponding author.

\bibliographystyle{aasjournal}
\bibliography{mydb}

\begin{thebibliography}{}
\expandafter\ifx\csname natexlab\endcsname\relax\def\natexlab#1{#1}\fi
\providecommand{\url}[1]{\href{#1}{#1}}
\providecommand{\dodoi}[1]{doi:~\href{http://doi.org/#1}{\nolinkurl{#1}}}
\providecommand{\doeprint}[1]{\href{http://ascl.net/#1}{\nolinkurl{http://ascl.net/#1}}}
\providecommand{\doarXiv}[1]{\href{https://arxiv.org/abs/#1}{\nolinkurl{https://arxiv.org/abs/#1}}}

\bibitem[{{Ashley} {et~al.}(2020){Ashley}, {Fox}, {Jenkins}, {Wakker}, {Bordoloi}, {Lockman}, {Savage}, \& {Karim}}]{2020ApJ...898..128A}
{Ashley}, T., {Fox}, A.~J., {Jenkins}, E.~B., {et~al.} 2020, \apj, 898, 128, \dodoi{10.3847/1538-4357/ab9ff8}

\bibitem[{{Baganoff} {et~al.}(2003){Baganoff}, {Maeda}, {Morris}, {Bautz}, {Brandt}, {Cui}, {Doty}, {Feigelson}, {Garmire}, {Pravdo}, {Ricker}, \& {Townsley}}]{2003ApJ...591..891B}
{Baganoff}, F.~K., {Maeda}, Y., {Morris}, M., {et~al.} 2003, \apj, 591, 891, \dodoi{10.1086/375145}

\bibitem[{Banda-Barrag{\'a}n {et~al.}(2016)Banda-Barrag{\'a}n, Parkin, Federrath, Crocker, \& Bicknell}]{BandaBarragan2016}
Banda-Barrag{\'a}n, W.~E., Parkin, E.~R., Federrath, C., Crocker, R.~M., \& Bicknell, G.~V. 2016, \mnras, 455, 1309, \dodoi{10.1093/mnras/stv2405}

\bibitem[{Barnes {et~al.}(2017)Barnes, Longmore, Battersby, Bally, Kruijssen, Henshaw, \& Walker}]{Barnes2017}
Barnes, A.~T., Longmore, S.~N., Battersby, C., {et~al.} 2017, \mnras, 469, 2263, \dodoi{10.1093/mnras/stx941}

\bibitem[{{Beck}(2013)}]{2013lsmf.book..215B}
{Beck}, R. 2013, in Large-Scale Magnetic Fields in the Universe, ed. R.~{Beck}, A.~{Balogh}, A.~{Bykov}, R.~A. {Treumann}, \& L.~{Widrow}, Vol.~39, 215--230, \dodoi{10.1007/978-1-4614-5728-2_8}

\bibitem[{Bordoloi {et~al.}(2017)Bordoloi, Fox, Lockman, Wakker, Jenkins, Savage, Hernandez, Tumlinson, Bland-Hawthorn, \& Kim}]{Bordoloi2017}
Bordoloi, R., Fox, A.~J., Lockman, F.~J., {et~al.} 2017, \apj, 834, 191, \dodoi{10.3847/1538-4357/834/2/191}

\bibitem[{{Butler} {et~al.}(2023){Butler}, {van der Werf}, {Topkaras}, {Rybak}, {Venemans}, {Walter}, \& {Decarli}}]{2023ApJ...944..134B}
{Butler}, K.~M., {van der Werf}, P.~P., {Topkaras}, T., {et~al.} 2023, \apj, 944, 134, \dodoi{10.3847/1538-4357/acad03}

\bibitem[{{Carretti} {et~al.}(2013){Carretti}, {Crocker}, {Staveley-Smith}, {Haverkorn}, {Purcell}, {Gaensler}, {Bernardi}, {Kesteven}, \& {Poppi}}]{2013Natur.493...66C}
{Carretti}, E., {Crocker}, R.~M., {Staveley-Smith}, L., {et~al.} 2013, \nat, 493, 66, \dodoi{10.1038/nature11734}

\bibitem[{{Cerri} {et~al.}(2017){Cerri}, {Gaggero}, {Vittino}, {Evoli}, \& {Grasso}}]{2017JCAP...10..019C}
{Cerri}, S.~S., {Gaggero}, D., {Vittino}, A., {Evoli}, C., \& {Grasso}, D. 2017, \jcap, 2017, 019, \dodoi{10.1088/1475-7516/2017/10/019}

\bibitem[{{Collins} {et~al.}(2004){Collins}, {Shull}, \& {Giroux}}]{2004ApJ...605..216C}
{Collins}, J.~A., {Shull}, J.~M., \& {Giroux}, M.~L. 2004, \apj, 605, 216, \dodoi{10.1086/382269}

\bibitem[{{Collins} {et~al.}(2005){Collins}, {Shull}, \& {Giroux}}]{2005ApJ...623..196C}
---. 2005, \apj, 623, 196, \dodoi{10.1086/428566}

\bibitem[{{Cooper} {et~al.}(2008){Cooper}, {Bicknell}, {Sutherland}, \& {Bland-Hawthorn}}]{2008ApJ...674..157C}
{Cooper}, J.~L., {Bicknell}, G.~V., {Sutherland}, R.~S., \& {Bland-Hawthorn}, J. 2008, \apj, 674, 157, \dodoi{10.1086/524918}

\bibitem[{{Crocker} \& {Aharonian}(2011)}]{2011PhRvL.106j1102C}
{Crocker}, R.~M., \& {Aharonian}, F. 2011, \prl, 106, 101102, \dodoi{10.1103/PhysRevLett.106.101102}

\bibitem[{Crocker {et~al.}(2015)Crocker, Bicknell, Taylor, \& Carretti}]{Crocker2015}
Crocker, R.~M., Bicknell, G.~V., Taylor, A.~M., \& Carretti, E. 2015, \apj, 808, 107, \dodoi{10.1088/0004-637X/808/2/107}

\bibitem[{{Di Teodoro} {et~al.}(2020){Di Teodoro}, {McClure-Griffiths}, {Lockman}, \& {Armillotta}}]{2020Natur.584..364D}
{Di Teodoro}, E.~M., {McClure-Griffiths}, N.~M., {Lockman}, F.~J., \& {Armillotta}, L. 2020, \nat, 584, 364, \dodoi{10.1038/s41586-020-2595-z}

\bibitem[{{Di Teodoro} {et~al.}(2018){Di Teodoro}, {McClure-Griffiths}, {Lockman}, {Denbo}, {Endsley}, {Ford}, \& {Harrington}}]{2018ApJ...855...33D}
{Di Teodoro}, E.~M., {McClure-Griffiths}, N.~M., {Lockman}, F.~J., {et~al.} 2018, \apj, 855, 33, \dodoi{10.3847/1538-4357/aaad6a}

\bibitem[{{Fabian}(2012)}]{2012ARA&A..50..455F}
{Fabian}, A.~C. 2012, \araa, 50, 455, \dodoi{10.1146/annurev-astro-081811-125521}

\bibitem[{{Ferri{\`e}re}(2009)}]{Ferriere2009}
{Ferri{\`e}re}, K. 2009, \aap, 505, 1183, \dodoi{10.1051/0004-6361/200912617}

\bibitem[{{Finkbeiner}(2004)}]{2004ApJ...614..186F}
{Finkbeiner}, D.~P. 2004, \apj, 614, 186, \dodoi{10.1086/423482}

\bibitem[{{Fox} {et~al.}(2015){Fox}, {Bordoloi}, {Savage}, {Lockman}, {Jenkins}, {Wakker}, {Bland-Hawthorn}, {Hernandez}, {Kim}, {Benjamin}, {Bowen}, \& {Tumlinson}}]{2015ApJ...799L...7F}
{Fox}, A.~J., {Bordoloi}, R., {Savage}, B.~D., {et~al.} 2015, \apjl, 799, L7, \dodoi{10.1088/2041-8205/799/1/L7}

\bibitem[{{Fujita} {et~al.}(2013){Fujita}, {Ohira}, \& {Yamazaki}}]{Fujita2013}
{Fujita}, Y., {Ohira}, Y., \& {Yamazaki}, R. 2013, \apjl, 775, L20, \dodoi{10.1088/2041-8205/775/1/L20}

\bibitem[{Fujita {et~al.}(2014)Fujita, Ohira, \& Yamazaki}]{Fujita2014}
Fujita, Y., Ohira, Y., \& Yamazaki, R. 2014, \apj, 789, 67, \dodoi{10.1088/0004-637X/789/1/67}

\bibitem[{Guo \& Mathews(2012)}]{Guo2012}
Guo, F., \& Mathews, W.~G. 2012, \apj, 756, 181, \dodoi{10.1088/0004-637X/756/2/181}

\bibitem[{{Haywood} {et~al.}(2016){Haywood}, {Lehnert}, {Di Matteo}, {Snaith}, {Schultheis}, {Katz}, \& {G{\'o}mez}}]{2016A&A...589A..66H}
{Haywood}, M., {Lehnert}, M.~D., {Di Matteo}, P., {et~al.} 2016, \aap, 589, A66, \dodoi{10.1051/0004-6361/201527567}

\bibitem[{{Heckman} \& {Best}(2014)}]{2014ARA&A..52..589H}
{Heckman}, T.~M., \& {Best}, P.~N. 2014, \araa, 52, 589, \dodoi{10.1146/annurev-astro-081913-035722}

\bibitem[{{Heckman} \& {Thompson}(2017)}]{2017hsn..book.2431H}
{Heckman}, T.~M., \& {Thompson}, T.~A. 2017, {Handbook of Supernovae}, ed. A.~W. {Alsabti} \& P.~{Murdin} (Springer International Publishing), 2431, \dodoi{10.1007/978-3-319-21846-5_23}

\bibitem[{{Heywood} {et~al.}(2019){Heywood}, {Camilo}, {Cotton}, {Yusef-Zadeh}, {Abbott}, {Adam}, {Aldera}, {Bauermeister}, {Booth}, {Botha}, {Botha}, {Brederode}, {Brits}, {Buchner}, {Burger}, {Chalmers}, {Cheetham}, {de Villiers}, {Dikgale-Mahlakoana}, {du Toit}, {Esterhuyse}, {Fanaroff}, {Foley}, {Fourie}, {Gamatham}, {Goedhart}, {Gounden}, {Hlakola}, {Hoek}, {Hokwana}, {Horn}, {Horrell}, {Hugo}, {Isaacson}, {Jonas}, {Jordaan}, {Joubert}, {J{\'o}zsa}, {Julie}, {Kapp}, {Kenyon}, {Kotz{\'e}}, {Kriel}, {Kusel}, {Lehmensiek}, {Liebenberg}, {Loots}, {Lord}, {Lunsky}, {Macfarlane}, {Magnus}, {Magozore}, {Mahgoub}, {Main}, {Malan}, {Malgas}, {Manley}, {Maree}, {Merry}, {Millenaar}, {Mnyandu}, {Moeng}, {Monama}, {Mphego}, {New}, {Ngcebetsha}, {Oozeer}, {Otto}, {Passmoor}, {Patel}, {Peens-Hough}, {Perkins}, {Ratcliffe}, {Renil}, {Rust}, {Salie}, {Schwardt}, {Serylak}, {Siebrits}, {Sirothia}, {Smirnov}, {Sofeya}, {Swart}, {Tasse}, {Taylor}, {Theron}, {Thorat}, {Tiplady}, {Tshongweni}, {van Balla}, {van der Byl},
  {van der Merwe}, {van Dyk}, {Van Rooyen}, {Van Tonder}, {Van Wyk}, {Wallace}, {Welz}, \& {Williams}}]{2019Natur.573..235H}
{Heywood}, I., {Camilo}, F., {Cotton}, W.~D., {et~al.} 2019, \nat, 573, 235, \dodoi{10.1038/s41586-019-1532-5}

\bibitem[{{Kataoka} {et~al.}(2013){Kataoka}, {Tahara}, {Totani}, {Sofue}, {Stawarz}, {Takahashi}, {Takeuchi}, {Tsunemi}, {Kimura}, {Takei}, {Cheung}, {Inoue}, \& {Nakamori}}]{2013ApJ...779...57K}
{Kataoka}, J., {Tahara}, M., {Totani}, T., {et~al.} 2013, \apj, 779, 57, \dodoi{10.1088/0004-637X/779/1/57}

\bibitem[{{Katsuda} {et~al.}(2018){Katsuda}, {Takiwaki}, {Tominaga}, {Moriya}, \& {Nakamura}}]{2018ApJ...863..127K}
{Katsuda}, S., {Takiwaki}, T., {Tominaga}, N., {Moriya}, T.~J., \& {Nakamura}, K. 2018, \apj, 863, 127, \dodoi{10.3847/1538-4357/aad2d8}

\bibitem[{Kroupa(2001)}]{Kroupa2001}
Kroupa, P. 2001, \mnras, 322, 231, \dodoi{10.1046/j.1365-8711.2001.04022.x}

\bibitem[{{Krumholz} \& {Kruijssen}(2015)}]{2015MNRAS.453..739K}
{Krumholz}, M.~R., \& {Kruijssen}, J.~M.~D. 2015, \mnras, 453, 739, \dodoi{10.1093/mnras/stv1670}

\bibitem[{{Leahy} \& {Williams}(2017)}]{Leahy2017a}
{Leahy}, D.~A., \& {Williams}, J.~E. 2017, \aj, 153, 239, \dodoi{10.3847/1538-3881/aa6af6}

\bibitem[{{Lockman} {et~al.}(2020){Lockman}, {Di Teodoro}, \& {McClure-Griffiths}}]{2020ApJ...888...51L}
{Lockman}, F.~J., {Di Teodoro}, E.~M., \& {McClure-Griffiths}, N.~M. 2020, \apj, 888, 51, \dodoi{10.3847/1538-4357/ab55d8}

\bibitem[{{McClure-Griffiths} {et~al.}(2013){McClure-Griffiths}, {Green}, {Hill}, {Lockman}, {Dickey}, {Gaensler}, \& {Green}}]{2013ApJ...770L...4M}
{McClure-Griffiths}, N.~M., {Green}, J.~A., {Hill}, A.~S., {et~al.} 2013, \apjl, 770, L4, \dodoi{10.1088/2041-8205/770/1/L4}

\bibitem[{{Michtchenko} \& {Barros}(2023)}]{2023A&A...680A..40M}
{Michtchenko}, T.~A., \& {Barros}, D.~A. 2023, \aap, 680, A40, \dodoi{10.1051/0004-6361/202347223}

\bibitem[{{Mignone} {et~al.}(2007){Mignone}, {Bodo}, {Massaglia}, {Matsakos}, {Tesileanu}, {Zanni}, \& {Ferrari}}]{Mignone2007}
{Mignone}, A., {Bodo}, G., {Massaglia}, S., {et~al.} 2007, \apjs, 170, 228, \dodoi{10.1086/513316}

\bibitem[{{Mignone} {et~al.}(2012){Mignone}, {Zanni}, {Tzeferacos}, {van Straalen}, {Colella}, \& {Bodo}}]{Mignone2012}
{Mignone}, A., {Zanni}, C., {Tzeferacos}, P., {et~al.} 2012, \apjs, 198, 7, \dodoi{10.1088/0067-0049/198/1/7}

\bibitem[{{Miyamoto} \& {Nagai}(1975)}]{1975PASJ...27..533M}
{Miyamoto}, M., \& {Nagai}, R. 1975, \pasj, 27, 533

\bibitem[{Mou {et~al.}(2014)Mou, Yuan, Bu, Sun, \& Su}]{Mou2014}
Mou, G., Yuan, F., Bu, D., Sun, M., \& Su, M. 2014, \apj, 790, 109, \dodoi{10.1088/0004-637X/790/2/109}

\bibitem[{Mou {et~al.}(2015)Mou, Yuan, Gan, \& Sun}]{Mou2015}
Mou, G., Yuan, F., Gan, Z., \& Sun, M. 2015, \apj, 811, 37, \dodoi{10.1088/0004-637X/811/1/37}

\bibitem[{{Mou} {et~al.}(2023){Mou}, {Sun}, {Fang}, {Wang}, {Zhang}, {Yuan}, {Sofue}, {Wang}, \& {He}}]{2023NatCo..14..781M}
{Mou}, G., {Sun}, D., {Fang}, T., {et~al.} 2023, Nature Communications, 14, 781, \dodoi{10.1038/s41467-023-36478-0}

\bibitem[{{Naab} \& {Ostriker}(2017)}]{2017ARA&A..55...59N}
{Naab}, T., \& {Ostriker}, J.~P. 2017, \araa, 55, 59, \dodoi{10.1146/annurev-astro-081913-040019}

\bibitem[{Nogueras-Lara {et~al.}(2020)Nogueras-Lara, Sch{\"o}del, Gallego-Calvente, Gallego-Cano, Shahzamanian, Dong, Neumayer, Hilker, Najarro, Nishiyama, Feldmeier-Krause, Girard, \& Cassisi}]{NoguerasLara2019}
Nogueras-Lara, F., Sch{\"o}del, R., Gallego-Calvente, A.~T., {et~al.} 2020, Nature Astronomy, 4, 377, \dodoi{10.1038/s41550-019-0967-9}

\bibitem[{{Planck Collaboration} {et~al.}(2013){Planck Collaboration}, {Ade}, {Aghanim}, {Arnaud}, {Ashdown}, {Atrio-Barandela}, {Aumont}, {Baccigalupi}, {Balbi}, {Banday}, {Barreiro}, {Bartlett}, {Battaner}, {Benabed}, {Beno{\^\i}t}, {Bernard}, {Bersanelli}, {Bonaldi}, {Bond}, {Borrill}, {Bouchet}, {Burigana}, {Cabella}, {Cardoso}, {Catalano}, {Cay{\'o}n}, {Chary}, {Chiang}, {Christensen}, {Clements}, {Colombo}, {Coulais}, {Crill}, {Cuttaia}, {Danese}, {D'Arcangelo}, {Davis}, {de Bernardis}, {de Gasperis}, {de Rosa}, {de Zotti}, {Delabrouille}, {Dickinson}, {Diego}, {Dobler}, {Dole}, {Donzelli}, {Dor{\'e}}, {D{\"o}rl}, {Douspis}, {Dupac}, {Efstathiou}, {En{\ss}lin}, {Eriksen}, {Finelli}, {Forni}, {Frailis}, {Franceschi}, {Galeotta}, {Ganga}, {Giard}, {Giardino}, {Gonz{\'a}lez-Nuevo}, {G{\'o}rski}, {Gratton}, {Gregorio}, {Gruppuso}, {Hansen}, {Harrison}, {Helou}, {Henrot-Versill{\'e}}, {Hern{\'a}ndez-Monteagudo}, {Hildebrandt}, {Hivon}, {Hobson}, {Holmes}, {Hornstrup}, {Hovest}, {Huffenberger}, {Jaffe},
  {Jagemann}, {Jewell}, {Jones}, {Juvela}, {Keih{\"a}nen}, {Knoche}, {Knox}, {Kunz}, {Kurki-Suonio}, {Lagache}, {L{\"a}hteenm{\"a}ki}, {Lamarre}, {Lasenby}, {Lawrence}, {Leach}, {Leonardi}, {Lilje}, {Linden-V{\o}rnle}, {L{\'o}pez-Caniego}, {Lubin}, {Mac{\'\i}as-P{\'e}rez}, {Maffei}, {Maino}, {Mandolesi}, {Maris}, {Marshall}, {Martin}, {Mart{\'\i}nez-Gonz{\'a}lez}, {Masi}, {Massardi}, {Matarrese}, {Matthai}, {Mazzotta}, {Meinhold}, {Melchiorri}, {Mendes}, {Mennella}, {Mitra}, {Moneti}, {Montier}, {Morgante}, {Munshi}, {Murphy}, {Naselsky}, {Natoli}, {N{\o}rgaard-Nielsen}, {Noviello}, {Novikov}, {Novikov}, {Osborne}, {Pajot}, {Paladini}, {Paoletti}, {Partridge}, {Pearson}, {Perdereau}, {Perrotta}, {Piacentini}, {Piat}, {Pierpaoli}, {Pietrobon}, {Plaszczynski}, {Pointecouteau}, {Polenta}, {Ponthieu}, {Popa}, {Poutanen}, {Pratt}, {Prunet}, {Puget}, {Rachen}, {Rebolo}, {Reinecke}, {Renault}, {Ricciardi}, {Riller}, {Ristorcelli}, {Rocha}, {Rosset}, {Rubi{\~n}o-Mart{\'\i}n}, {Rusholme}, {Sandri}, {Savini},
  {Schaefer}, {Scott}, {Smoot}, {Spencer}, {Stivoli}, {Sudiwala}, {Suur-Uski}, {Sygnet}, {Tauber}, {Terenzi}, {Toffolatti}, {Tomasi}, {Tristram}, {T{\"u}rler}, {Umana}, {Valenziano}, {Van Tent}, {Vielva}, {Villa}, {Vittorio}, {Wade}, {Wandelt}, {White}, {Yvon}, {Zacchei}, \& {Zonca}}]{2013A&A...554A.139P}
{Planck Collaboration}, {Ade}, P.~A.~R., {Aghanim}, N., {et~al.} 2013, \aap, 554, A139, \dodoi{10.1051/0004-6361/201220271}

\bibitem[{{Ponti} {et~al.}(2019){Ponti}, {Hofmann}, {Churazov}, {Morris}, {Haberl}, {Nandra}, {Terrier}, {Clavel}, \& {Goldwurm}}]{2019Natur.567..347P}
{Ponti}, G., {Hofmann}, F., {Churazov}, E., {et~al.} 2019, \nat, 567, 347, \dodoi{10.1038/s41586-019-1009-6}

\bibitem[{{Predehl} {et~al.}(2020){Predehl}, {Sunyaev}, {Becker}, {Brunner}, {Burenin}, {Bykov}, {Cherepashchuk}, {Chugai}, {Churazov}, {Doroshenko}, {Eismont}, {Freyberg}, {Gilfanov}, {Haberl}, {Khabibullin}, {Krivonos}, {Maitra}, {Medvedev}, {Merloni}, {Nandra}, {Nazarov}, {Pavlinsky}, {Ponti}, {Sanders}, {Sasaki}, {Sazonov}, {Strong}, \& {Wilms}}]{Predehl2020}
{Predehl}, P., {Sunyaev}, R.~A., {Becker}, W., {et~al.} 2020, \nat, 588, 227, \dodoi{10.1038/s41586-020-2979-0}

\bibitem[{{Roberts-Borsani} {et~al.}(2020){Roberts-Borsani}, {Saintonge}, {Masters}, \& {Stark}}]{2020MNRAS.493.3081R}
{Roberts-Borsani}, G.~W., {Saintonge}, A., {Masters}, K.~L., \& {Stark}, D.~V. 2020, \mnras, 493, 3081, \dodoi{10.1093/mnras/staa464}

\bibitem[{{Sarkar}(2024)}]{2024A&ARv..32....1S}
{Sarkar}, K.~C. 2024, \aapr, 32, 1, \dodoi{10.1007/s00159-024-00152-1}

\bibitem[{Sarkar {et~al.}(2015)Sarkar, Nath, \& Sharma}]{Sarkar2015}
Sarkar, K.~C., Nath, B.~B., \& Sharma, P. 2015, \mnras, 453, 3827, \dodoi{10.1093/mnras/stv1806}

\bibitem[{{Sofue} \& {Handa}(1984)}]{1984Natur.310..568S}
{Sofue}, Y., \& {Handa}, T. 1984, \nat, 310, 568, \dodoi{10.1038/310568a0}

\bibitem[{{Sormani} {et~al.}(2020){Sormani}, {Tress}, {Glover}, {Klessen}, {Battersby}, {Clark}, {Hatchfield}, \& {Smith}}]{2020MNRAS.497.5024S}
{Sormani}, M.~C., {Tress}, R.~G., {Glover}, S. C.~O., {et~al.} 2020, \mnras, 497, 5024, \dodoi{10.1093/mnras/staa1999}

\bibitem[{{Spilker} {et~al.}(2018){Spilker}, {Aravena}, {B{\'e}thermin}, {Chapman}, {Chen}, {Cunningham}, {De Breuck}, {Dong}, {Gonzalez}, {Hayward}, {Hezaveh}, {Litke}, {Ma}, {Malkan}, {Marrone}, {Miller}, {Morningstar}, {Narayanan}, {Phadke}, {Sreevani}, {Stark}, {Vieira}, \& {Wei{\ss}}}]{2018Sci...361.1016S}
{Spilker}, J.~S., {Aravena}, M., {B{\'e}thermin}, M., {et~al.} 2018, Science, 361, 1016, \dodoi{10.1126/science.aap8900}

\bibitem[{{Spilker} {et~al.}(2020){Spilker}, {Aravena}, {Phadke}, {B{\'e}thermin}, {Chapman}, {Dong}, {Gonzalez}, {Hayward}, {Hezaveh}, {Litke}, {Malkan}, {Marrone}, {Narayanan}, {Reuter}, {Vieira}, \& {Wei{\ss}}}]{2020ApJ...905...86S}
{Spilker}, J.~S., {Aravena}, M., {Phadke}, K.~A., {et~al.} 2020, \apj, 905, 86, \dodoi{10.3847/1538-4357/abc4e6}

\bibitem[{{Stuber} {et~al.}(2021){Stuber}, {Saito}, {Schinnerer}, {Emsellem}, {Querejeta}, {Williams}, {Barnes}, {Bigiel}, {Blanc}, {Dale}, {Grasha}, {Klessen}, {Kruijssen}, {Leroy}, {Meidt}, {Pan}, {Rosolowsky}, {Schruba}, {Sun}, \& {Usero}}]{2021A&A...653A.172S}
{Stuber}, S.~K., {Saito}, T., {Schinnerer}, E., {et~al.} 2021, \aap, 653, A172, \dodoi{10.1051/0004-6361/202141093}

\bibitem[{{Su} {et~al.}(2010){Su}, {Slatyer}, \& {Finkbeiner}}]{2010ApJ...724.1044S}
{Su}, M., {Slatyer}, T.~R., \& {Finkbeiner}, D.~P. 2010, \apj, 724, 1044, \dodoi{10.1088/0004-637X/724/2/1044}

\bibitem[{{Sukhbold} {et~al.}(2016){Sukhbold}, {Ertl}, {Woosley}, {Brown}, \& {Janka}}]{Sukhbold2016}
{Sukhbold}, T., {Ertl}, T., {Woosley}, S.~E., {Brown}, J.~M., \& {Janka}, H.-T. 2016, \apj, 821, 38, \dodoi{10.3847/0004-637X/821/1/38}

\bibitem[{{Truelove} \& {McKee}(1999)}]{Truelove1999}
{Truelove}, J.~K., \& {McKee}, C.~F. 1999, \apjs, 120, 299, \dodoi{10.1086/313176}

\bibitem[{{Veena} {et~al.}(2024){Veena}, {Kim}, {S{\'a}nchez-Monge}, {Schilke}, {Menten}, {Fuller}, {Sormani}, {Wyrowski}, {Banda-Barrag{\'a}n}, {Riquelme}, {Tarr{\'\i}o}, \& {de Vicente}}]{2024A&A...689A.121V}
{Veena}, V.~S., {Kim}, W.~J., {S{\'a}nchez-Monge}, {\'A}., {et~al.} 2024, \aap, 689, A121, \dodoi{10.1051/0004-6361/202450902}

\bibitem[{{Yang} \& {Ruszkowski}(2017)}]{Yang2017}
{Yang}, H.-Y.~K., \& {Ruszkowski}, M. 2017, \apj, 850, 2, \dodoi{10.3847/1538-4357/aa9434}

\bibitem[{{Yang} {et~al.}(2013){Yang}, {Ruszkowski}, \& {Zweibel}}]{2013MNRAS.436.2734Y}
{Yang}, H. Y.~K., {Ruszkowski}, M., \& {Zweibel}, E. 2013, \mnras, 436, 2734, \dodoi{10.1093/mnras/stt1772}

\bibitem[{{Yang} {et~al.}(2022){Yang}, {Ruszkowski}, \& {Zweibel}}]{2022NatAs...6..584Y}
{Yang}, H. Y.~K., {Ruszkowski}, M., \& {Zweibel}, E.~G. 2022, Nature Astronomy, 6, 584, \dodoi{10.1038/s41550-022-01618-x}

\bibitem[{{Zhang}(2018)}]{2018Galax...6..114Z}
{Zhang}, D. 2018, Galaxies, 6, 114, \dodoi{10.3390/galaxies6040114}

\bibitem[{{Zhang} {et~al.}(2017){Zhang}, {Thompson}, {Quataert}, \& {Murray}}]{2017MNRAS.468.4801Z}
{Zhang}, D., {Thompson}, T.~A., {Quataert}, E., \& {Murray}, N. 2017, \mnras, 468, 4801, \dodoi{10.1093/mnras/stx822}

\bibitem[{{Zhang} {et~al.}(2024){Zhang}, {Ponti}, {Carretti}, {Liu}, {Morris}, {Haverkorn}, {Locatelli}, {Zheng}, {Aharonian}, {Zhang}, {Zhang}, {Stel}, {Strong}, {Yeung}, \& {Merloni}}]{2024NatAs.tmp..228Z}
{Zhang}, H.-S., {Ponti}, G., {Carretti}, E., {et~al.} 2024, Nature Astronomy, \dodoi{10.1038/s41550-024-02362-0}

\bibitem[{{Zhang} \& {Li}(2024)}]{2024MNRAS.527.3418Z}
{Zhang}, M., \& {Li}, M. 2024, \mnras, 527, 3418, \dodoi{10.1093/mnras/stad3408}

\bibitem[{{Zhang} {et~al.}(2021){Zhang}, {Li}, \& {Morris}}]{2021ApJ...913...68Z}
{Zhang}, M., {Li}, Z., \& {Morris}, M.~R. 2021, \apj, 913, 68, \dodoi{10.3847/1538-4357/abf927}

\bibitem[{Zhang \& Guo(2020)}]{Zhang2020}
Zhang, R., \& Guo, F. 2020, \apj, 894, 117, \dodoi{10.3847/1538-4357/ab8bd0}

\bibitem[{Zubovas {et~al.}(2011)Zubovas, King, \& Nayakshin}]{Zubovas2011}
Zubovas, K., King, A.~R., \& Nayakshin, S. 2011, \mnras, 415, L21, \dodoi{10.1111/j.1745-3933.2011.01070.x}

\bibitem[{{Zubovas} \& {Nayakshin}(2012)}]{2012MNRAS.424..666Z}
{Zubovas}, K., \& {Nayakshin}, S. 2012, \mnras, 424, 666, \dodoi{10.1111/j.1365-2966.2012.21250.x}

\end{thebibliography}

\appendix

\section{The gravitational potential model}\label{sec:grav}
The simulation applies a new Galactic gravitational potential model \citep{2023A&A...680A..40M}, which includes the contribution of stellar disk (thin and thick disk), gaseous components (HI and H$_2$), bulge and halo.
Each of components is a superposition of three Miyamoto-Nagai discs \citep{1975PASJ...27..533M}, $\Phi^1_{\mathrm{MN}}(R,z)$, $\Phi^2_{\mathrm{MN}}(R,z)$ and $\Phi^3_{\mathrm{MN}}(R,z)$. We adopted the 1st-, 2nd-,and 3rd-order expressions for the potential of the MN-discs,
\begin{eqnarray}
    \begin{aligned}
    & \Phi_{\mathrm{MN}}^1(R, z)=\frac{-G M}{\sqrt{R^2+(a+\zeta)^2}} \\
    & \Phi_{\mathrm{MN}}^2(R, z)=\Phi_{\mathrm{MN}}^1(R, z)-\frac{G M a(a+\zeta)}{\left[R^2+(a+\zeta)^2\right]^{3 / 2}} \\
    & \Phi_{\mathrm{MN}}^3(R, z)=\Phi_{\mathrm{MN}}^2(R, z)+\frac{G M}{3} \times \frac{a^2\left[R^2-2(a+\zeta)^2\right]}{\left[R^2+(a+\zeta)^2\right]^{5 / 2}}
    \end{aligned}
\end{eqnarray}
in which, $R = \sqrt{x^2 +y^2}$, $\zeta = \sqrt{z^2 +b^2}$, with $M, a, b$ being the mass, the radial scale length and vertical scale length.

\begin{figure}
\includegraphics[width=0.5\textwidth,keepaspectratio]{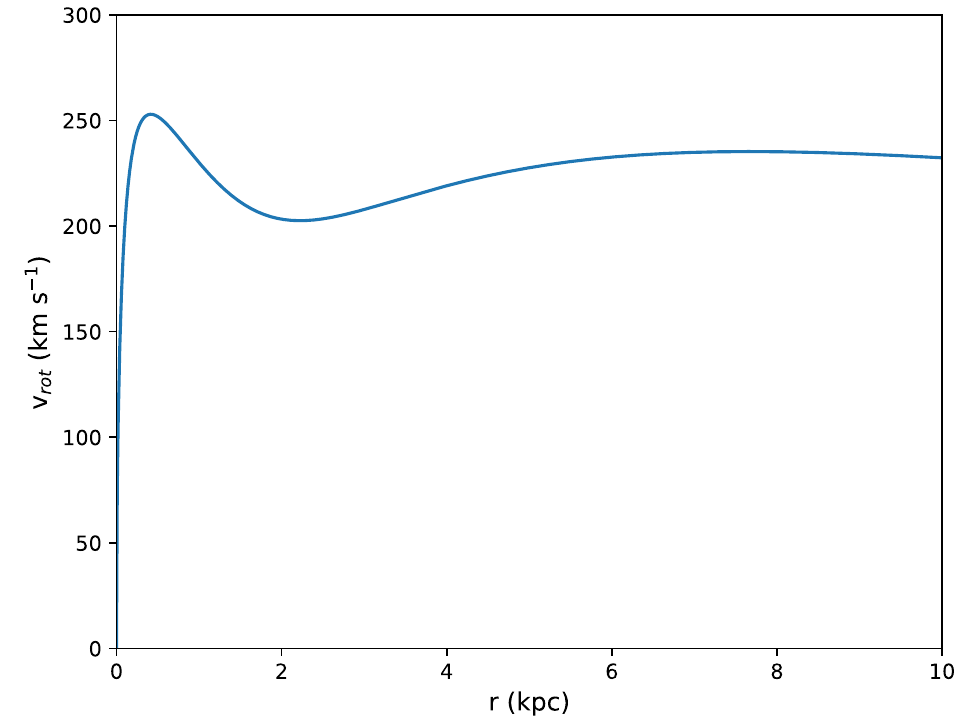}
\caption{The rotation curve at the Galactic plane derived from the gravitational potential used in the simulations.
\label{fig:grav}}
\end{figure}

The potential of stellar thin/thick disks are
\begin{equation}
    \Phi_{\text {thin }}(R, z)=\sum_{i=1}^3 \Phi_{\mathrm{MN} i}^3(R, z),
\end{equation}
\begin{equation}
    \Phi_{\text {thick }}(R, z)=\sum_{i=1}^3 \Phi_{\mathrm{MN} i}^1(R, z).
\end{equation}
The potential of gaseous HI and H$_2$ disks are respectively
\begin{equation}
    \Phi_{\text {HI }}(R, z)=\sum_{i=1}^3 \Phi_{\mathrm{MN} i}^2(R, z),
\end{equation}
\begin{equation}
    \Phi_{\text {H$_2$}}(R, z)=\sum_{i=1}^3 \Phi_{\mathrm{MN} i}^3(R, z).
\end{equation}
The potential of bulge and halo are respectively
\begin{equation}
\Phi_{\mathrm{b}}(R, z)=\frac{-G M}{\sqrt{R^2+z^2}+a},
\end{equation}
\begin{equation}
\Phi_{\mathrm{h}}(R, z)=\frac{v_{\mathrm{h}}^2}{2} \ln \left(R^2+z^2+r_{\mathrm{h}}^2\right),
\end{equation}
where $v_\mathrm{h}$ is the circular velocity at large $R$ and $r_\mathrm{h}$ is the core radius.
The parameters used in the simulation follow the work of \citet{2023A&A...680A..40M}. The total potential used in the simulation is the sum of the six components, then the rotation curve at the Galactic disk can be derived (see Figure~\ref{fig:grav}).

\section{Cooling function}\label{sec:cf}
\begin{figure}
\includegraphics[width=0.5\textwidth,keepaspectratio]{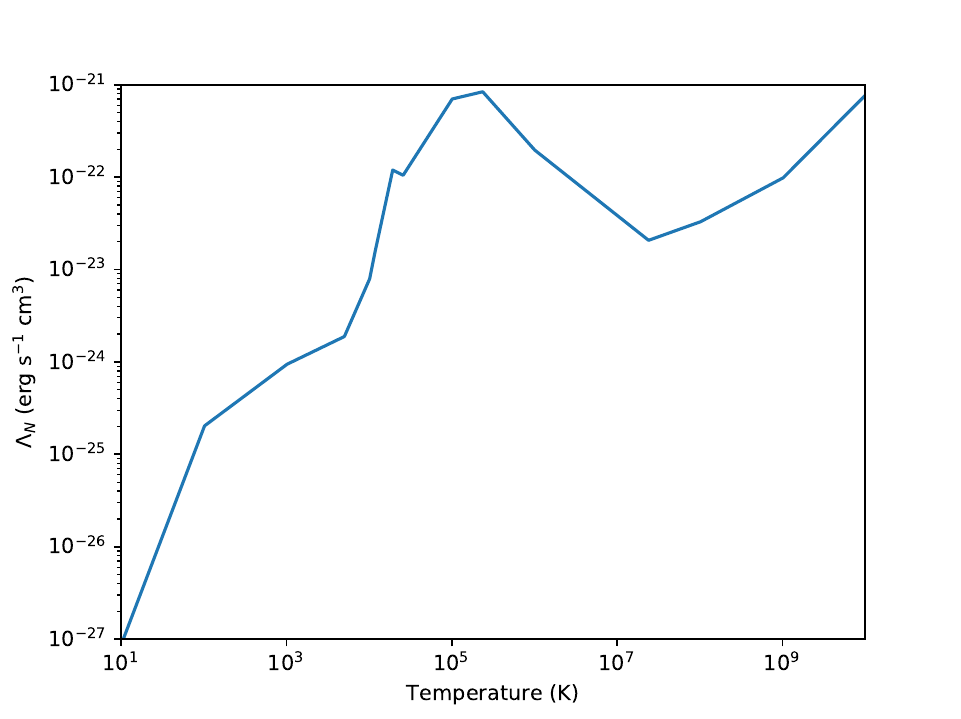}
\caption{The piecewise cooling curve used in the simulations.
\label{fig:cool}}
\end{figure}

The cooling process can significantly influence the evolution of HVMCs, but an accurate tabulated cooling function will spend much more computational resource.
Therefore, we in the simulations adopt a piecewise cooling function (see Figure~\ref{fig:cool}), which can roughly describe the cooling function.

\section{Conversion to Local Standard Rest Velocity}\label{sec:lsr}
Similar to the kinematic wind model of \citet{2018ApJ...855...33D}, we apply a left-handed cylindrical coordinate system centered on the Galactic center.
The $z$-axis is set to be perpendicular to the Galactic disk (north as positive), the $y$-axis to run along decreasing Galactic longitude, and the $x$-axis to be parallel to the line-of-sight (the observer at the positive side).
The local standard rest velocity of HVCs is estimated by
\begin{equation}
    \begin{aligned}
    V_{\mathrm{LSR}}= & \left(V_\theta \frac{R_{\odot}}{R}-V_{\odot}\right) \sin (l) \cos (b) +V_z \sin (b)-V_R \cos (\ell+\theta) \cos (b)
    \end{aligned}
\end{equation}
where $V_\theta$ is the tangential velocity, $R_{\odot}$ the distance from solar system, $V_{\odot}$ the solar velocity, $l$ the Galactic longitude, $b$ the Galactic latitude, $V_z$ the vertical velocity, $V_R$ the radial velocity, $\theta = \mathrm{sign}(y)\arccos(x/\sqrt{x^2+y^2}) - \pi(\mathrm{sign}(y)-1)$, with a function of 'sign' used to choose the sign of y.

Because the tangential velocity is also considered in the simulation, therefore, it cannot be simplified by a purely radial case, different from the model used by \citet{2018ApJ...855...33D}.

\end{document}